\documentclass[sn-nature,iicol]{sn-jnl}

\usepackage{xcolor} 

\usepackage{verbatim}
\usepackage{subcaption}
\usepackage{hyperref}

\usepackage{graphicx}%
\usepackage{multirow}%
\usepackage{amsmath,amssymb,amsfonts}%
\usepackage{amsthm}%
\usepackage{mathrsfs}%
\usepackage[title]{appendix}%
\usepackage{xcolor}%
\usepackage{textcomp}%
\usepackage{manyfoot}%
\usepackage{booktabs}%
\usepackage{algorithm}%
\usepackage{algorithmicx}%
\usepackage{algpseudocode}%
\usepackage{listings}%

\theoremstyle{thmstyleone}%
\theoremstyle{thmstyletwo}%

\theoremstyle{thmstylethree}%

\begin{document}

\title[Enhancing Yeast Cell Tracking with a Time-Symmetric Deep Learning Approach]{Enhancing Yeast Cell Tracking with a Time-Symmetric Deep Learning Approach}

\author*[1]{\fnm{Gergely} \sur{Szabó}}\email{szabo.gergely@itk.ppke.hu}

\author[2]{\fnm{Paolo} \sur{Bonaiuti}}\email{paolo.bonaiuti@ifom.eu}

\author[1,2]{\fnm{Andrea} \sur{Ciliberto}}\email{andrea.ciliberto@ifom.eu}

\author*[1]{\fnm{András} \sur{Horváth}}\email{horvath.andras@itk.ppke.hu}

\affil*[1]{\orgdiv{ITK}, \orgname{PPCU}, \orgaddress{\street{ Pr\'ater st. 50/A}, \city{Budapest}, \postcode{1083}, \country{Hungary}}}

\affil[2]{\orgname{IFOM}, \orgaddress{\street{Via Adamello, 16}, \city{Milan}, \postcode{20139}, \country{Italy}}}

\abstract{
The accurate tracking of live cells using video microscopy recordings remains a challenging task for popular state-of-the-art image processing based object tracking methods. In recent years, several existing and new applications have attempted to integrate deep-learning based frameworks for this task, but most of them still heavily rely on consecutive frame based tracking embedded in their architecture or other premises that hinder generalized learning. To address this issue, we aimed to develop a new deep-learning based tracking method that relies solely on the assumption that cells can be tracked based on their spatio-temporal neighborhood, without restricting it to consecutive frames. The proposed method has the additional benefit that the motion patterns of the cells can be learned completely by the predictor without any prior assumptions, and it has the potential to handle a large number of video frames with heavy artifacts. The efficacy of the proposed method is demonstrated through biologically motivated validation strategies and compared against multiple state-of-the-art cell tracking methods on budding yeast recordings and simulated samples.
}

\keywords{Cell Tracking, Deep Learning, Image Processing, Time-Symmetric Approach, Video Microscopy}

\maketitle

\section{Introduction}

For several decades, tracking of solid objects in video recordings has been a partially solved problem. Classical image processing-based pattern matching approaches, such as Scale-Invariant Feature Transform (SIFT) \cite{SIFT} combined with Random Sample Consensus (RANSAC) \cite{RANSAC}, have the potential to track objects with fixed shapes invariantly to shift, scale, rotation, and partially to lighting conditions. Furthermore the extension of such methods with variants of the Kalman filter \cite{KalmanFilter}\cite{UnscentedKalman} can introduce object velocity or other degrees of freedom into the estimation, which can substantially reduce the position estimation noise, particularly when the introduced additional parameters tend to be continuous. However, these methods base their estimation on prior assumptions, which greatly reduces their capabilities when the assumptions do not hold true. These limitations are especially apparent in live cell tracking on videomicroscopy recordings, as cells tend to change shape and morphological description rapidly, unpredictably, and in a highly nonlinear manner. Additionally, cells can grow and divide with time, have no real velocity, and often exhibit unpredictable behavior as a living, non-rigid object. Furthermore, cells on the same recordings tend to be very similar, making the task even more challenging. Therefore, the most promising solution to address these issues is using machine learning, where the feature importances and other estimates are learned by the model and depend mainly on the distribution of the training dataset.

In recent years, several cell tracking software and libraries have switched to machine-learning-based solutions, mainly utilizing convolutional neural networks (CNNs), as they generally perform well on image processing tasks. Some of these tools, such as CellTracker \cite{hu2021celltracker}, Usiigaci \cite{tsai2019usiigaci}, and YeaZ \cite{dietler2020convolutional}, separate segmentation and tracking and use deep learning only for high-quality segmentation. However, they use classical methods, such as the Kalman filter, bipartite graph matching, or other methods for tracking which ultimately rely on a rigid metric system. In contrast, other tools, such as CellTrack R-CNN \cite{chen2021celltrack}, and Ilastik \cite{berg2019ilastik}, adopt deep learning or other forms of machine learning for tracking either in an end-to-end manner or as a separate segmentation followed by tracking. This approach shows advantages both theoretically and empirically, as demonstrated by the often superior performance of machine learning based and other data-driven methods when sufficient amount of training data is available. \cite{vlah2022data}\cite{sun2017revisiting} However, to the best of our knowledge, all of these tools rely on a frame-by-frame identity matching approach, utilizing only consecutive frames for tracking. This approach is a major simplification and disregards a large amount of useful information that could be present in the temporal neighborhood of the given video frames. 

Therefore, we developed a complete tracking pipeline that separates segmentation and tracking because end-to-end training with our method was unfeasible. Our method uses deep learning for both stages utilizing the temporal neighborhood of consecutive frames for prediction, makes skipping of frames possible up to the size of the temporal neighborhood making the tracking substantially more robust and uses metrics only between different predictions for the same cell on the same frame minimizing the introduction of methodical assumptions via the choice of the similarity metric. The most similar method in the literature that we know of is DeepTrack \cite{katariya2022deeptrack}, which was recently developed for tracking cars. However, while this method utilizes the local temporal neighborhood via encoding objects using Temporal Convolutional Networks (TCNs), the network architecture and the metric system for object matching is substantially different, making our solution architecturally novel.

\subsection{Yeast Data Source}

The training and validation data used to generate the results in this paper are videos of budding yeast cells dividing in a microfluidic device.  Cells are \emph{S. Cerevisiae} wild-type-like, with W303 genetic background. Yeast is a unicellular eukaryote, which shares many essential genes and phenotypes with multi-cellular eukaryotes, such as mammals. On the other hand, its relatively small genome ($\sim 12 \times 10^6$ base pairs), short doubling time and easy genetic tools allow to perform experiments in a short time and with limited costs. Such experiments would be very costly or even impossible to do in higher eukaryotes. Not surprisingly, many discoveries performed in higher eukaryotes were inspired by original studies performed in yeast.

In the experiments, cells were grown overnight at 30~$^\circ$C in complete YPD medium and synchronized by the use of the pheromone $\alpha$-factor. Imaging started when they were released from this stimulus. Cells were free to grow and duplicate, trapped in the microfluidic device to prevent them from moving in the field of view. Images were acquired every 3~minutes for 3.5~hours using a DeltaVision Elite imaging system equipped with a phase-contrast objective. The field of view is a square of 111.1 $\mu$m size, resulting in a 512x512 pixels image. The duration of the cell-cycle of the analysed cells is~$\sim 1.5$ hours, while their size is~$\sim 5 \mu$m.

The segmentation and tracking models were trained on a dataset comprising of 314 movies, and for validation purposes during model design, 35 additional movies were used. These movies were acquired using the same microscope, objective, and image size, but with varying durations and time-lapses. On average, each movie contained ~$35.8$ frames and ~$20.2$ individual cells to be tracked. The ground truth labels were initially generated using Phylocell \cite{charvin2021phylocell}, but they were subsequently corrected and curated by multiple experts. The training and validation sample numbers of the segmentation model were independent of model parameters, while for the tracking model, the model parameter tracking range ($TR$) substantially influenced the number of available samples, as shown in Table \ref{tab:sample_nums}. Further details on the model parameter $TR$ will be discussed in Section \ref{sec:local_tracking}. For final testing and evaluation of the tools and model parameters, 5 additional independent recordings were used, containing a total of 4629 total number of identified cell instances.

\begin{table}[]
\resizebox{\columnwidth}{!}{
\begin{tabular}{llll}
\toprule
\textbf{\begin{tabular}[c]{@{}l@{}}Tracking \\ Range\end{tabular}} & \textbf{\begin{tabular}[c]{@{}l@{}}Valid Local \\ Tracks\end{tabular}} & \textbf{Training Samples} & \textbf{Validation Samples} \\
\midrule
1 & 99019 & 88993 & 10026 \\
2 & 64659 & 58059 & 6600 \\
3 & 47940 & 43018 & 4922 \\
4 & 37347 & 33487 & 3860 \\
5 & 29888 & 26769 & 3119 \\
6 & 24373 & 21803 & 2570 \\
7 & 19931 & 17819 & 2112 \\
8 & 16278 & 14533 & 1745 \\
9 & 13301 & 11847 & 1454 \\
10 & 10915 & 9709 & 1206 \\
\bottomrule
\end{tabular}
}
\caption{Number of valid local tracking samples for model training in function of tracking range ($TR$), and the disribution of training and validation samples.}
\label{tab:sample_nums}
\end{table}

\subsection{Comparative Datasets} \label{sec:comp_datasets_descr}

While our architecture was initially designed for the segmentation and tracking of yeast cells, we also claim that it is capable of generalization to diverse data types, including those featuring faster-moving objects, given an adequate amount of training data. Additionally, in contrast to numerous other methods, our tracking architecture seamlessly integrates both morphological and kinematic information, making it entirely data-driven and retrainable without requiring any substantial architectural changes.

To support these claims, we aimed to train and evaluate it using microscopic cell recordings of other cell types. However, we found that such data is not readily available in the required quantity and quality. Our architecture demands extensive training data due to its size and complexity, as a trade-off for tracking quality and stability, which poses a challenge compared to simpler tracking solutions. For instance, the Cell Tracking Challenge \cite{mavska2023cell} or the CTMC-v1 dataset \cite{anjum2020ctmc} could have been suitable for evaluation purposes. However, they lack the necessary intra-class variance required for successful training of complex models without severe overfitting on the training data. This issue was also noted in the publication of I. E. Toubal (2023) \cite{toubal2023ensemble}.

To address this issue, we synthetically created multiple toy datasets comprising objects with various morphological and motility patterns. To introduce a level of difficulty and realism to the tracking tasks, each simulated recording includes a randomized background composed of 10 Gaussian density functions, with a maximum relative brightness level of 0.39, and a similar framewise-random noise with a maximum relative brightness level of 0.078.

\subsubsection*{Synthetic Arrows}

The objects designated for tracking are triangular arrows, with a maximum size difference ratio of 6, an average speed equivalent to 40\% of the mean object size, uncorrelated with the size of the individual object, allowing for speeds even greater than the object size. Additionally, these objects have a maximum angular rotation of 10° for each frame, allowing for object crossing and thus occlusion, and random uniform grayscale values exceeding a brightness level of 0.39. Furthermore, the objects may also expand from one frame to the next by 1.6-10.0\% of their original size, with a probability of 0.33 at each time step.

\subsubsection*{Synthetic Amoeboids}

This object tracking task is similar to "Synthetic Arrows". However, the objects to be tracked are more complex amoeboids with semi-random initial shapes and shape changes from one frame to the next. Both the initial shapes of these objects and the iterative shape changes are generated using Perlin noise \cite{perlin1985image} in a 50-point polar coordinate system with an octave value of 6, a persistence value of 0.5, and a lacunarity value of 2.0. Furthermore, the shape changes are governed in a Gaussian manner based on the distance from the object centroid to avoid complete filling of the object space or the disappearance of an object due to random chance. This approach creates amoeboid-like objects with randomized boundaries allowing for highly concave contours and even the short-term splitting of an object in extreme cases. Additionally, while the individual shape changes are gradual from one frame to the next, the objects can be nearly unrecognizable over larger temporal distances. Lastly, unlike the "Synthetic Arrows", the objects perform perfectly rigid collisions with no velocity loss, making occlusion impossible. The main reason for this choice is that the following other datasets featuring synthetic amoeboids would require a realistic simulation of occluded object light transference, which would be mandatory but exceedingly challenging if object occlusions were allowed. Furthermore, object occlusions are generally less important in microscopy compared to macroscopic cameras, due to the focal plane specificity of microscopes.

\subsubsection*{Synthetic Amoeboids-PC}

The aim of this dataset is to simulate object tracking on phase contrast microscopy recordings with objects displaying substantially different morphological and motion characteristics compared to yeast cells. The behavior of the objects in this dataset is identical to that of "Synthetic Amoeboids." However, instead of displaying objects with uniform grayscale color, the Canny edge detector \cite{canny1986computational} is applied to the binary object mask, followed by a Gaussian blur with a large kernel size of 51 pixels, mimicking phase contrast microscopy visual characteristics in a simplistic manner.

\subsubsection*{Synthetic Amoeboids-PCC}

The objects in this dataset behave similarly to those in "Synthetic Amoeboids-PC," with the only notable difference being that the objects undergo non-rigid collisions, resulting in a 10\% velocity loss for each collision involving both objects. This seemingly minor change leads to the clumping of objects, making the tracking task simultaneously easier and more challenging. Object identification and segmentation become substantially more difficult due to the clumping, while the reduced momentum simplifies assignment. Additionally, a small force is applied to each object towards the center of the field of view. Although this force does not visibly alter object motility, it ensures that the object clump is likely to form within the field of view.

\subsubsection*{Synthetic Amoeboids-PCCA}

This dataset does not exhibit any differences in terms of object behavior compared to "Synthetic Amoeboids-PCC." Instead, each recording is augmented with a highly disruptive artifact, which poses challenges for both object detection and tracking, even for the human eye. These artifacts consist of 100 white lines placed in a uniform random manner between the edges of the field of view, obscuring both the objects and the background. The artifact patterns generated in this way are static but unique to each recording, rendering it impossible for models to learn their positions.

\subsection{Segmentation}

In case of tracking pipelines not trained in an end-to-end manner, segmentation is commonly employed as a preliminary stage before tracking. The reason being that the detection of the objects is mandatory for tracking, and detection via segmentation can provide essential information about the objects present, simplifying and improving the manageability of the tracking task. In the context of cell tracking, high-quality segmentation of cells is also important from a biological perspective as it offers valuable information regarding the size and shape of cells, as well as serving as a quality check for the tracking process.

From both the target and solution architecture perspectives, segmentation can be classified into semantic segmentation and instance segmentation. In the realm of CNN-based image processing, semantic segmentation is typically considered the easier task and can be accomplished using relatively simpler network architectures like SegNet \cite{SegNet}, U-Net \cite{Unet}, and others. However, accurately separating objects of the same category after semantic segmentation can be highly challenging and may necessitate the utilization of classical image processing methods, which we previously discussed as being less desirable. On the other hand, recent advancements in deep learning, particularly the introduction of Mask R-CNN-like architectures \cite{MaskRCNN}, have made it possible to achieve instance segmentation solely based on deep learning techniques. Although these architectures tend to be more complex and rigid due to the combination of convolutional, fully connected, and potentially other network connection types, they are highly valuable when it is essential to separate instances of objects belonging to the same category.

In our approach, we employ a distinct step of Mask R-CNN-based instance segmentation prior to tracking, grounded in the previously described arguments. This step serves two purposes: to provide accurate segmentation results and to serve as initialization for the tracker. Additionally, the tracking method itself partially relies on segmentation too. However, in this context, the objective is to separate a single cell from the background and other cells, rendering semantic segmentation adequate for the task. In subsequent sections we give a detailed architectural description for of both components of our pipeline.

\subsection{Tracking}

While deep-learning based segmentation has gained popularity in modern cell tracking software, tracking often relies on comparing consecutive frames using metrics like Euclidean distance or Intersection over Union (IOU). In contrast, more advanced but still rigid methods, such as variants of the Kalman filter, consider motility patterns along with position and shape for tracking. Following such frame-to-frame similarity measures, various techniques, including variants of the Hungarian method \cite{HungarianMethod} can be employed for optimal assignment without repetition.

However, these methods are sensitive to several hyper-parameters, artifacts, and unusual cases. Even with perfect segmentation and correct assignment, frame-to-frame metrics are not expected to yield a perfect match unless the object moves in a completely predictable way, with estimable process and measurement noises. \cite{UnscentedKalman} This sensitivity arises due to the movement, speed, shape changes, and other descriptors of the object, as well as environmental factors such as lighting conditions. Consequently, manual fine-tuning of parameters on a per-recording basis, or even adopting different parameters within a single record, is often necessary.

Machine-learning based techniques offer an alternative to address these challenges, as they can learn the movement patterns of the tracked object and adapt to environmental changes when provided with diverse training data. Artificial data augmentation can also be used to enhance data variety or simulate specific anomalies. However, in popular cell tracking software, these solutions are usually implemented on a frame-to-frame level, potentially leading to sensitivity to unexpected temporally local anomalies, such as changes in lighting conditions or administration of drugs that alter cell movement patterns or morphology.

Instead we propose a multi-frame based assignment method that predicts the position of a marked cell on several consecutive frames using a state-of-the-art segmentation network. The assignment of local tracks is then carried out using the Hungarian method, permitting multiple frame skipping on a hierarchical level that decreases with increasing temporal distance.

\section{Architectural Design and Features}

\subsection{Instance Segmentation}

The emergence of Mask R-CNNs and similar architectures, coupled with the availability of libraries and software packages utilizing it, has greatly simplified instance segmentation for various applications. This holds true for segmenting living cells as well, although quality can greatly vary depending on the cell type and the training data at hand. Therefore, instead of focusing on developing a novel architecture, we trained an instance segmentation model using the Detectron2 \cite{Detectron2} environment. To improve the performance and robustness of the model, we employed multiple data augmentation techniques during the training process.

The specific model we chose was a Mask R-CNN architecture with ResNet-50 feature pyramid backbone pretrained on the COCO instance segmentation dataset \cite{COCO} with 128 ROI heads for the single "cell" object type. The model was trained through 120,000 iterations with a base learning rate of 0.0025, minibatch size of 4 and otherwise the deafult parameters of the Detectron2 environment. For artificial data augmentation, we employed various transformation options provided by the Detectron2 environment and the Albumentations library \cite{buslaev2020albumentations}. The chosen transformations and their respective parameters are summarized in Table \ref{tab:ins_seg_transformations}. The transformations and their parameters were determined in a qualitative manner to ensure that the resulting outcomes closely resemble biologically plausible samples, while maximizing variance. However, conducting an extensive data-specific parameter search could potentially yield even better results.

\begin{table}[]
\centering
\resizebox{\columnwidth}{!}{
\begin{tabular}{ll}
\toprule
\multicolumn{2}{c}{Detectron2 transformations} \\
\midrule
Transformation & Parameters \\
\midrule
Random Brightness & Intensity Min = 0.7, Intensity Max = 1.3, P = 0.2 \\
Random Contrast & Intensity Min = 0.7, Intensity Max = 1.3, P = 0.2 \\
Random Flip & P = 0.1 \\
Random Extent & Intensity Min = 0.7, Intensity Max = 1.3, No Shift, P = 0.2 \\
\bottomrule
\end{tabular}
}

\vspace{0.5cm}

\resizebox{\columnwidth}{!}{
\begin{tabular}{ll}
\toprule
\multicolumn{2}{c}{Albumentations transformations} \\
\midrule
Transformation & Parameters \\
\midrule
Grid Distortion & Num Steps = 10, Distort Limit = 0.2, P = 0.2 \\
Elastic Transformation & Alpha = 991, Sigma = 8, Alpha Affine = 50, P = 0.2 \\
\bottomrule
\end{tabular}
}
\caption{Data augmentation transformation parameters for Detectron2 and Albumentations-based instance segmentation training}
\label{tab:ins_seg_transformations}
\end{table}

\begin{figure}[]
    \centering
    \begin{subfigure}[b]{0.32\columnwidth}
        \centering
        \includegraphics[width=\linewidth]{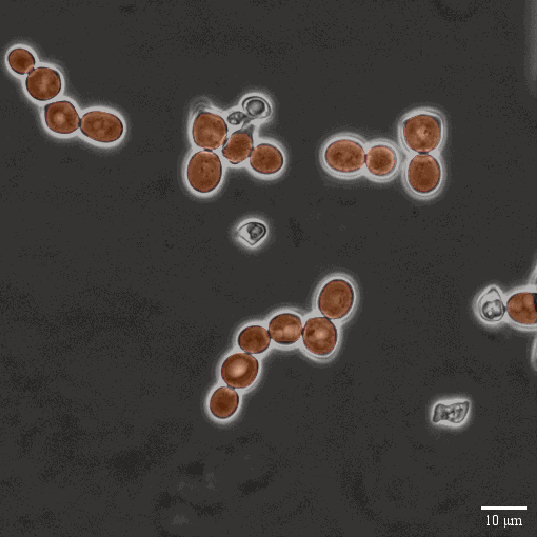}
    \end{subfigure}%
    \hfill
    \begin{subfigure}[b]{0.32\columnwidth}
        \centering
        \includegraphics[width=\linewidth]{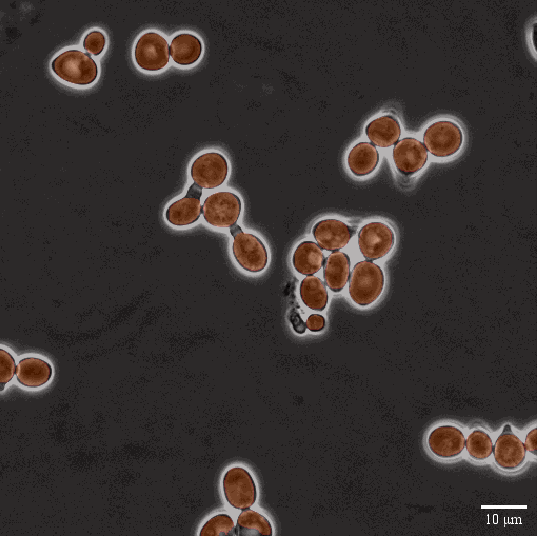}
    \end{subfigure}%
    \hfill
    \begin{subfigure}[b]{0.32\columnwidth}
        \centering
        \includegraphics[width=\linewidth]{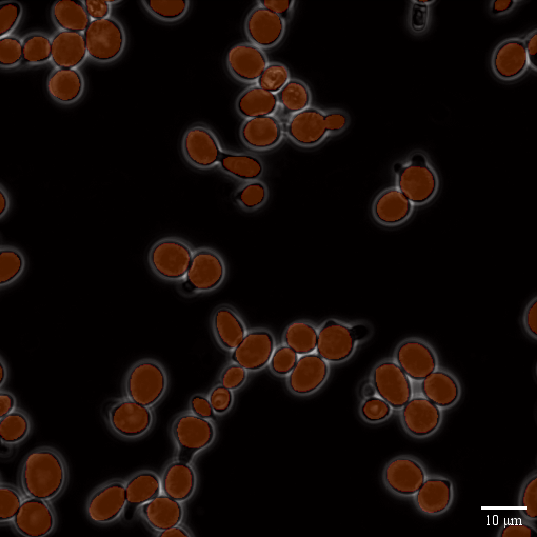}
    \end{subfigure}
    \caption{A showcase of robust segmentation results, depicting accurate cell detection and segmentation under varying lighting conditions, while effectively avoiding detection of obviously dead cells.}
    \label{fig:segmentation_results}
\end{figure}

The trained instance segmentation model showed exceptional performance in both object detection and shape recognition. Therefore we decided to not only use its results as inputs for the subsequent tracking method in the pipeline but also utilize them as the final segmentation output. This decision was made in contrast to the option of using segmentation estimates of the tracking method. Examples of such instance segmentation results can be seen in Figure \ref{fig:segmentation_results}. Using these initial instance segmentation predictions as segmentation instances in tracking effectively combines the strengths of both stages of the pipeline: the segmentator module produces more accurate segmentations, while the tracker connects them.

\subsection{Tracking}

Our method is based on the concept that, although the direction of time is important in cell tracking, certain morphological changes such as cell growth generally occur in the positive temporal direction. Therefore, having information about the future position of the cell can be advantageous for tracking. As a result, we have developed a time-symmetric tracking method that utilizes the local temporal neighborhood in both the negative and positive temporal directions. Based on this architecture, the model learns to capture the direction of time and temporally unidirectional changes from the data, instead of relying on a temporally biased architecture. Our approach draws notable inspiration from the 2019 study of X. Wang \cite{wang2019learning}. However, instead of using cycle consistency losses, we sought to create an architecture that remains invariant to the direction of time itself. After the predictions are generated by this architecture, the resulting local tracks can be optimally matched and linked together based on prediction information within their temporal overlaps.

\subsubsection*{Local Tracking}\label{sec:local_tracking}

The local tracking around a single frame is performed on all cells individually distinguished by the previous instance segmentation step via a neural network designed for semantic segmentation. The channel parameters of this network can be defined the tracking range ($TR$) parameter. The segmentator model input for a single cell instance consists of $(TR+1+TR)+1$ images as channels of the single input where $TR+1+TR$ images are raw video frames centered around the frame on which the given cell is to be tracked, while on the additional $1$ channel a single white dot marks the cell to be tracked at its centroid and the rest of the image is black. The determination of the centroid coordinates $(x_c, y_c)$ is based on the segmentation instance for the given cell using image moments. The equations are as follows:

\begin{equation}
\label{eq:centroid_x}
x_c = \frac{\sum\sum x \cdot I(x, y)}{\sum\sum I(x, y)}
\end{equation}
\begin{equation}
\label{eq:centroid_y}
y_c = \frac{ \sum\sum y \cdot I(x, y)}{\sum\sum I(x, y)}
\end{equation}

\begin{figure}[]
  \centering
  \includegraphics[width=\columnwidth]{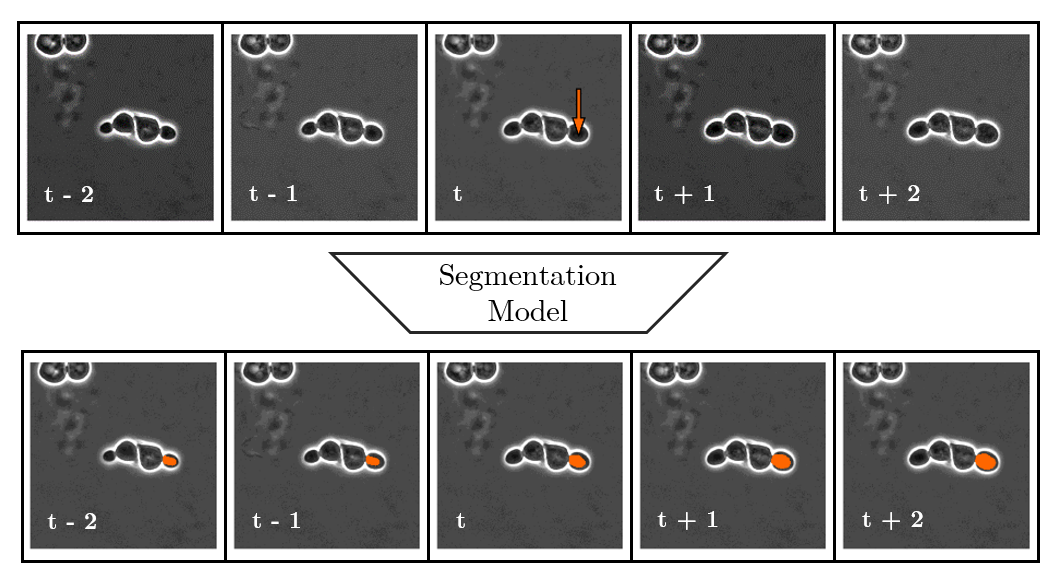}
  \caption{Depiction of input and output information structure of local tracking via segmentation with local tracking range ($TR$) = 2. Notice that while the left to right positive temporal direction of the data is recognizable from cell growth, and the model learns this during training, the architecture itself has no built in directional preference, resulting in a completely data-driven estimation for local tracking.}
  \label{fig:local_tracking_step}
\end{figure}

The target outputs for the local tracker consist of $n+1+n$ frames, containing the segmentation of the cell marked on the last channel of the input. An illustration of this local tracking architecture is presented in Figure \ref{fig:local_tracking_step}.

The reason for using the centroid of the segmentation and not the segmentation itself in the input is that according to our experiments if the segmentation information would be present there, the local tracker would not learn how to perform semantic segmentation. Instead it would copy the input segmentation to all outputs, providing false segmentations to all outputs except the middle one. However via using the centroid as a marker, the local tracker is forced to learn how to perform segmentation for the single marked cell on all temporal instances present, and thus it will be able to perform tracking via segmentation. Furthermore, if a cell instance is temporally close to the beginning or end of the recording, such that it is within n frames of the boundary, direct forward or backward tracking using a kernel size of n+1+n is not feasible. Nevertheless, the number of channels, which determines the local tracking distance, is a fixed parameter of the network architecture. To address this issue, we incorporate temporal padding by repeating the first or last frame. Additionally, we ensured that such instances are included during training to make the network capable of handling such edge cases and maintain its temporal invariance.

Based on empirical evidence, we found that among the tested semantic segmentation models, variants of the DeepLabV3+ \cite{deeplabv3plus2018} architecture from Pytorch Segmentation Models \cite{Iakubovskii:2019} yielded the best performance for this task. The models were trained through 20 epochs with a minibatch size of 10 in all experiments using stochastic gradient descent optimizer with cosine annealing learning rate scheduling having warm restarts during the first 15 epochs, and a gradual cooldown during the last 5 epochs. The model output had sigmoid activation with binary crossentropy loss.

Artificial data augmentation of the local tracking samples was performed using the torchvision transformations library \cite{torchvision}. For positional transformations random horizontal flip [p=0.5] and random affine [degrees=(-40, 40), scale=(0.7, 1.3), p=0.4] transformations were uniformly applied to the input and ground truth data. For color transformations color jitter [brightness=0.5, contrast=0.5, p=0.4] transformation was applied for each input frame individually to improve model robustness for unexpected lighting artefacts.

\subsubsection*{Global Tracking}

As the output of local tracking, a single cell instance marked by the video frame and its position is tracked forwards and backwards through $TR$ frames. This local tracking output can be obtained for each cell instance detected by the instance segmentation step, serving as the input for global tracking. Therefore the aim of global tracking is to match local tracks based on the semantic segmentation predictions in their overlapping areas, and chain them together based on ID matching creating full cell tracks based on a globally optimal consensus. 

In practice, at first, we obtain this global consensus on a $t$ to $t+1$ level, as usually the cells can be tracked from one frame to the next. However, if a cell instance did not receive any matches above the minimal requirement threshold, the matching can be repeated for larger temporal distances. This secondary matching is performed after all $t$ to $t+1$ matching is done, and only between the remaining candidates. Secondary matching is also performed hierarchically in terms of temporal difference meaning that temporally closer instances get matched first and only the remaining ones can get matched later with larger temporal differences. The maximal possible matching distance is $2TR$, but matches over $TR$ are unreliable as they are based on relatively few segmentation predictions, and only indirectly contain information about the segmentation of the central cell instances to be matched. Therefore we believe, using $TR$ as the maximal temporal distance between local track matching is logically the most sound choice, but in practice other choices could have minor benefits depending on the dataset. 

\begin{figure}[]
  \centering
  \includegraphics[width=\columnwidth]{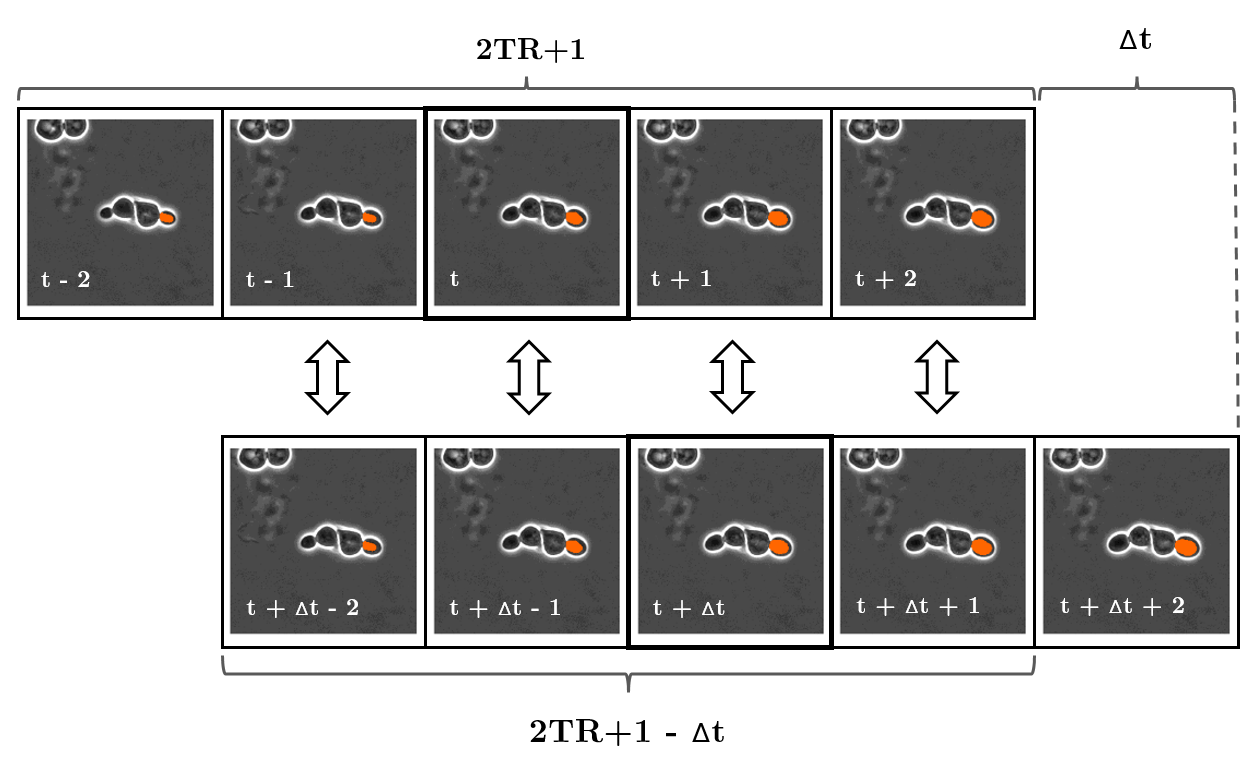}
  \caption{Schematic structure of the metric similarity measurement step between $2TR+1$ long local tracks of different cell instances on frames with a temporal distance of $\Delta t$. The solid lines indicate the central segmentation instances with a $\Delta t$ temporal distance to be matched, while the arrows indicate the similarity metric between the segmentation estimates for each time frame. Subsequently, the individual metric results are averaged to obtain a single measure of similarity.}
  \label{fig:assignment_step}
\end{figure}

Based on this methodology, the global consensus of ID matching can be turned into a one-to-one assignment problem for $N\times M$ candidates, where $N$ is the number of candidates in frame $t$ and $M$ is the number of candidates in frame $t+\Delta t$ where $1<\Delta t\le2TR$. Such assignment problems can be optimally solved by variants of the Hungarian method in polynomial time even for non-square matrices, but this requires a sound metric choice to measure the goodness of the candidate matches. In our case these candidate matches are the $2TR+1-\Delta t$ segmentation pairs in the overlapping temporal region of the local tracks to be compared. Therefore in an ideal scenario where perfect segmentations are achieved in local tracking, any metrics that solely compare segments within the same frame and calculate the average of these comparisons could be employed. This is because segmentation instances belonging to the same cell should result in a perfect match. However in practice, the mean of Intersection over Union (IOU) turned out to be the best metric choice as it takes both the positional and morphological differences of the segmentations into account, and gives 0 similarity for segmentations with no overlap regardless of the distance. This process of calculating the metric similarity measure based on local tracks is depicted in Figure \ref{fig:assignment_step}.

\begin{figure}[]
    \centering
    \begin{subfigure}[b]{0.48\columnwidth}
        \centering
        \includegraphics[width=\linewidth]{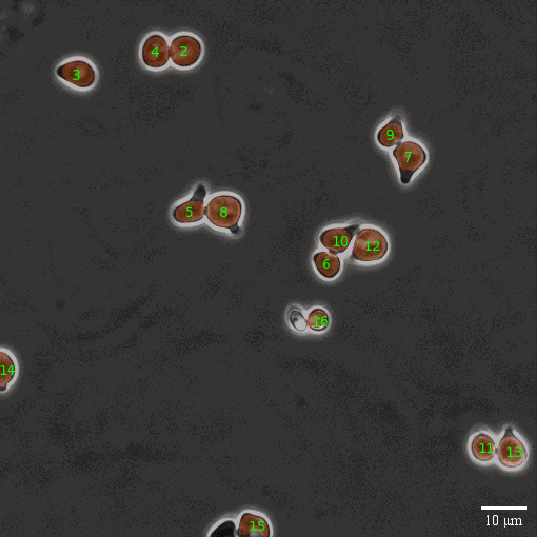}
    \end{subfigure}
     \begin{subfigure}[b]{0.48\columnwidth}
        \centering
        \includegraphics[width=\linewidth]{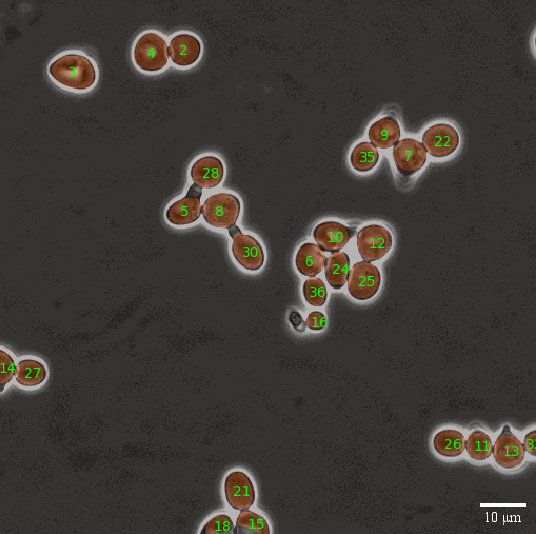}
    \end{subfigure}
    \caption{A side by side display of tracking results, demonstrating successful tracking of all cells in the same recording with a temporal difference of 8 frames. New cells were appropriately assigned new IDs while maintaining consistent tracking of existing cells.}
    \label{fig:tracking_results}
\end{figure}

\begin{figure}[]
  \centering
  \includegraphics[width=\columnwidth]{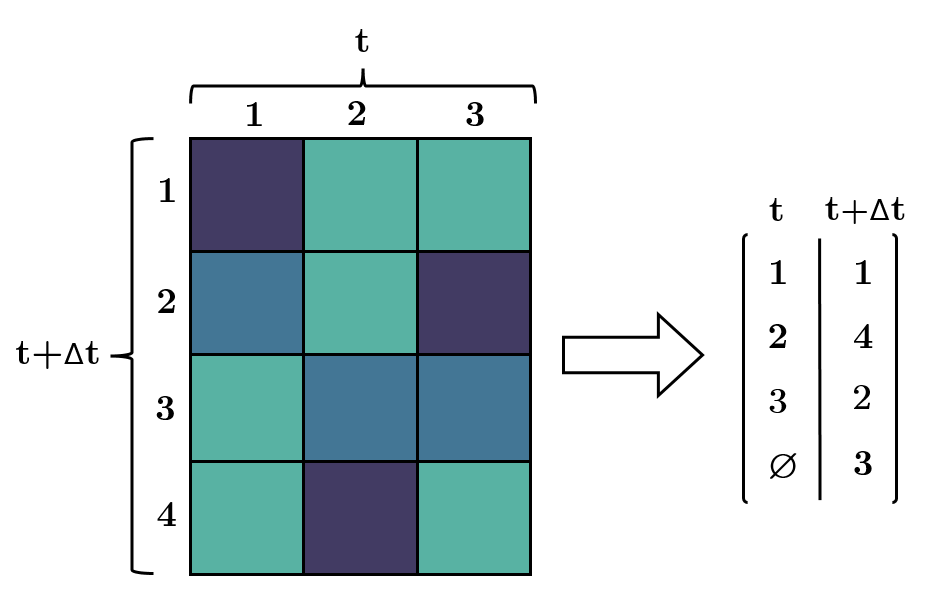}
  \caption{Illustration of the metric similarity based assignment step using the Hungarian method for non-square matrices. Newly unassigned cells, such as cell number 4 receive a new ID, while the previously assigned cells retain the ID of their corresponding previous instances.}
  \label{fig:hungarian_step}
\end{figure}

After computing the selected metric between all global tracks of frame $t$ and $t+\Delta t$, a threshold can be employed to establish a minimum required similarity based on the metric results. This step eliminates candidates that do not meet the desired level of similarity. Subsequently, the Hungarian method can be applied to determine the optimal global consensus for ID matching between frame $t$ and $t+\Delta t$, as depicted in Figure \ref{fig:hungarian_step}. Via performing these steps in the hierarchical approach as described earlier, complete global tracks are generated based on a global consensus of similarity, which solely relies on the prediction capabilities of the local tracker. An example of successful global tracking using this pipeline is presented in Figure \ref{fig:tracking_results}.

\subsubsection*{Skipped Instance Interpolation}

By employing the aforementioned methodology for global tracking, we can effectively address inconsistency errors in segmentation that commonly arise in realistic scenarios when $\Delta t > 1$. In such cases, although the resulting tracks will be complete, they may not be continuous, and segmentations for missed frames will be absent. To tackle this issue, we use a positional linear interpolation method, which is described by the following equations:

\begin{equation}
\label{eq:interpolation_dp}
\Delta c(x,y) = \frac{\begin{aligned}(t - t_{\text{last}})c_{\text{next}}(x,y) + \\ (t_{\text{next}} - t)c_{\text{last}}(x,y)\end{aligned}}{t_{\text{next}} - t_{\text{last}}}
\end{equation}
\begin{equation}
\label{eq:interpolation_S}
S(x,y) = S_{\text{last}}(x,y) + c(x,y)
\end{equation}

Here, $S(x,y)$ and $t$ represent the interpolated segmentation and its corresponding time, respectively. $S_{\text{last}}(x,y)$, $c_{\text{last}}(x,y)$, and $t_{\text{last}}$ denote the segmentation, centroid, and time of the last occurrence of the cell with the given ID, while $c_{\text{next}}(x,y)$ and $t_{\text{next}}$ represent the centroid and time of the next occurrence of the cell with the given ID. Although this method only shifts the segmentation from the last occurrence to the linearly interpolated position without altering its shape, in practice the results have proven satisfactory. Still, in the future, incorporating a method that linearly interpolates the shape of the segmentation by considering both the last and next occurrences could potentially yield minor improvements.

\subsubsection*{Ancestry Assingment}

Our architecture currently does not include an integrated solution for assigning newly born cells to their ancestors. However, the stability and continuity of the predicted cell tracks suggest the feasibility of implementing such a feature later if required. Initial tests show that a basic prediction method using Euclidean distance, combined with positional and morphological heuristics like maximal newborn cell size, can produce satisfactory results for ancestry assignment when the predicted tracks are unbroken. For example, a somewhat similar simple distance-based method was employed in the Cell Tracking Challenge submission by A. Arbelle (2021) \cite{mavska2023cell}\cite{arbellebgu}. However, for more complex samples, additional morphological factors may need consideration, such as connectedness, or machine learning-based solutions could be utilized as a separate module if an ample amount of training data is available. Additionally, there is the option of integrated object instance classification of division, akin to the approach described in the publication by I. E. Toubal (2023) \cite{toubal2023ensemble}. However, this would further increase the data intensity of architecture training and could introduce a potential weakness in the tracker module due to false positive cell division predictions dissecting some of the tracks.

\section{Evaluation}

For evaluation purposes, we opted to calculate F-scores based on binary decisions of correctness for segmentation and tracking, instead of utilizing continuous regression metrics. This approach offers a clear numerical evaluation of the capabilities of the compared tools and tool versions. F-scores for segmentation were computed using IOU, with a minimum similarity threshold of $0.5$. We selected this threshold as it is strict enough to assess the accuracy of correct cell detection while disregarding minor segmentation discrepancies that hold minimal biological significance. In the case of tracking evaluation, F-scores were also calculated based on the segmentation IOU, with the additional inclusion of ID matching. True positive values, also referred to as "links" were registered if the segmentations matched on the same frame, and the IDs matched consecutively for both the prediction and the ground truth. This tracking evaluation method of link matching is described in the work of M. Primet (2011) \cite{primet2011probabilistic}.

\subsection{Comparative Evaluation}

For comparative evaluation, we selected Phylocell \cite{charvin2021phylocell} \cite{fehrmann2013aging} and YeaZ \cite{padovani2022segmentation} as the two other cell tracking tools. The reason behind this choice was that the ground truth training and testing data were generated using Phylocell and were later manually corrected by experts. On the other hand YeaZ provides a fair comparison to our solution as it employs a similar segmentation pipeline and a somewhat comparable tracking pipeline, utilizing the Hungarian method. As the ground truth data excludes cells that eventually leave the microscope's field of view throughout the recording (from now on referred to as "border tracks"), we conducted two evaluation scenarios for the tools. In one scenario, border tracks were removed in post-processing from all predictions to ensure an unbiased comparison, while in the other scenario, the prediction results were left unaltered.

\begin{figure*}[]
    \centering
    \begin{subfigure}[b]{0.75\columnwidth}
        \centering
        \includegraphics[width=\linewidth]{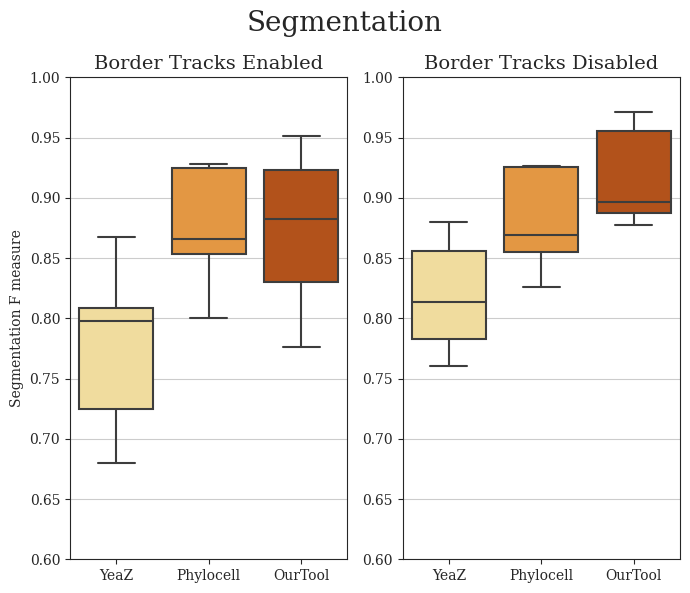}
    \end{subfigure}
     \begin{subfigure}[b]{0.75\columnwidth}
        \centering
        \includegraphics[width=\linewidth]{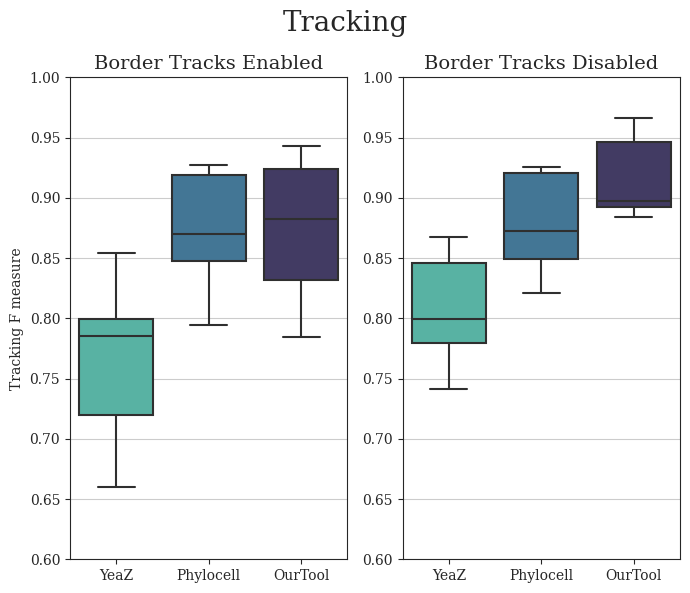}
    \end{subfigure}
    \caption{Comparative evaluation of segmentation and tracking performances of Phylocell, YeaZ and our pipeline based on F-measures of segmentation IOU and tracking link matches with enabled and disabled border track predictions. Disabled border tracks present an unbiased comparison between the tools, while enabled border tracks show results with the unaltered outputs.}
    \label{fig:comparative_eval}
\end{figure*}

The segmentation and tracking F-measure results are depicted in Figure \ref{fig:comparative_eval}. It is evident that our solution substantially surpasses both other tools in terms of segmentation and tracking quality, particularly in the more equitable scenario where the border tracks were eliminated. In the unaltered prediction case, the performance of Phylocell remained comparable to its prior results since the tool already automatically removes most border tracks. Conversely, YeaZ and our solution exhibited notably lower performance due to domain discrepancy between the predictions and ground truth data. Nevertheless, even in this case, our solution outperformed Phylocell in terms of expected F-score values, although with a substantially higher degree of variance on the lower side.

These results indicate the effectiveness of our method, but also suggest at a strong connection between segmentation and tracking quality due to the apparent correlation between metric results. To investigate further, we used Phylocell instance segmentation inputs for our tracker. In this case, the mean tracking F-score of our method was $0.868\pm0.015$, showing a negligible difference compared to the tracking F-score of Phylocell at $0.878\pm0.02$. While our tracking method did not surpass the performance of Phylocell on the measured samples, this may be attributed to a highly non-uniform segmentation error distribution of Phylocell, which leads to certain track predictions being mostly incorrect while others are mostly correct. Although our tracker is theoretically capable of correcting sparse errors in mostly correct tracks, it is unable to rectify tracks composed predominantly of faulty or missing instance predictions. To verify this, we conducted an experiment detailed in Section \ref{sec:tracking_robustness}, introducing segmentation errors in a more uniform manner.

\subsection{Hyperparameter Dependencies}

The parameters we considered to have a major impact on tracking performance for both local and global tracking are the local tracking range, the complexity of the local tracker model backbone, and the metric used for global consensus.

\begin{figure*}[]
    \centering
    \begin{subfigure}[b]{0.66\columnwidth}
        \centering
        \includegraphics[width=\linewidth]{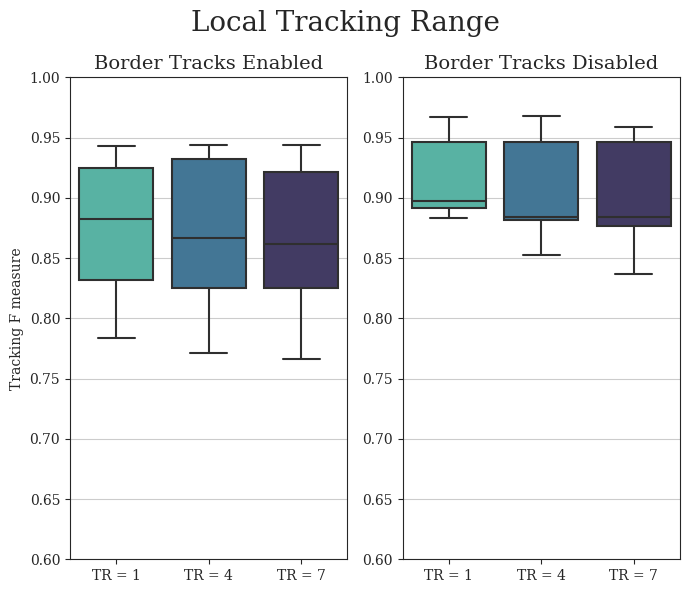}
    \end{subfigure}
    \begin{subfigure}[b]{0.66\columnwidth}
        \centering
        \includegraphics[width=\linewidth]{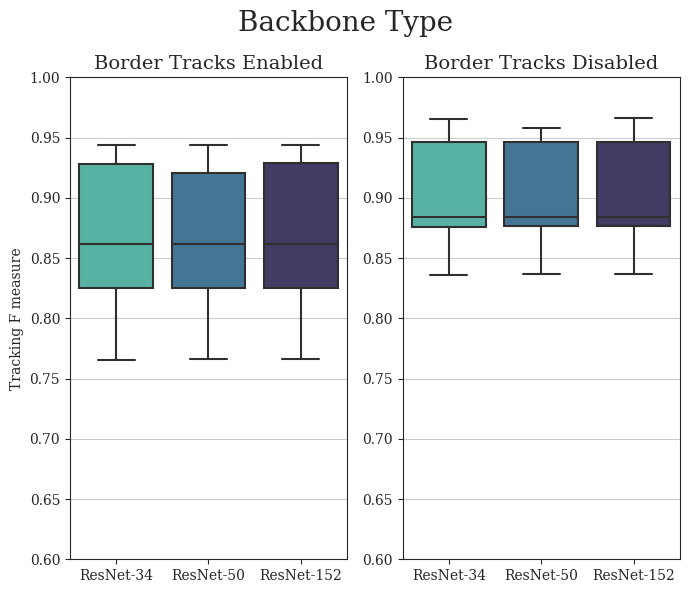}
    \end{subfigure}
    \begin{subfigure}[b]{0.66\columnwidth}
        \centering
        \includegraphics[width=\linewidth]{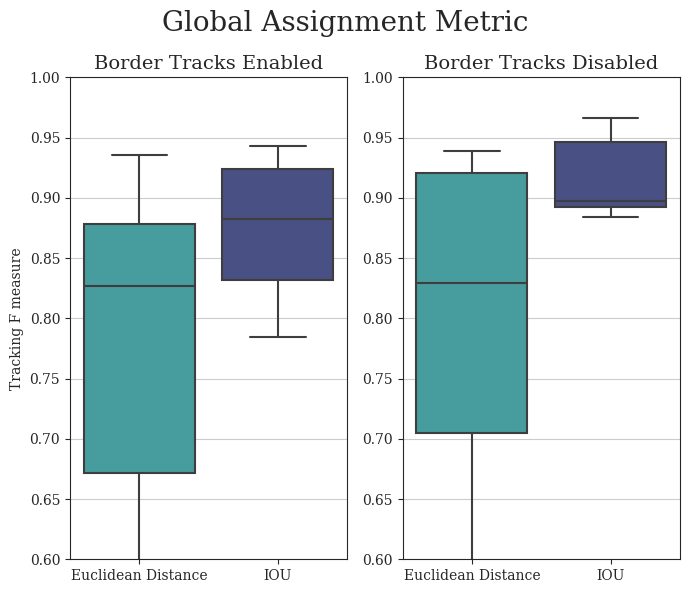}
    \end{subfigure}
    \caption{Comparative evaluation of tracking F-scores in function of local tracking ranges, backbone model complexities and global consensus metrics with enabled and disabled border track predictions. Disabled border tracks present an unbiased comparison between the parameters, while enabled border tracks show results with the unaltered outputs.}
    \label{fig:hyperparam_eval}
\end{figure*}

The differences in tracking F-scores resulting from these parameters are presented in Figure \ref{fig:hyperparam_eval}. Somewhat surprisingly, we observed minimal differences in performance based on the local tracking range and model complexity. Furthermore, smaller local tracking ranges exhibited marginally better performance, suggesting that the benefits of larger tracking kernels were outweighed by the increased model complexity, which slightly hindered training. On the other hand, these results also indicate that lightweight models with smaller local tracking ranges and simpler backbones are suitable for tracking predictions, resulting in substantially shorter inference times. As an additional benefit, while performing inference without GPU acceleration remains slower by orders of magnitude, it is more reasonable with such lightweight models, making initial pipeline testing more accessible. As anticipated, the use of IOU as the metric for global consensus substantially outperformed Euclidean distance, underscoring the importance of incorporating cell morphology in short-term tracking.

\subsection{Tracking Robustness}\label{sec:tracking_robustness}

To comprehensively demonstrate the tracking stabilization and segmentation interpolation capabilities of our architecture, and to show that tracking quality is not solely reliant on initial instance segmentation quality, we conducted an ablation study. In this study, we removed every 1:15 (Noise 1:15) and 1:5 (Noise 1:5) segmentation instances in a uniform random manner before tracking. Furthermore, to showcase the potential advantage of longer local tracking ranges, we created a scenario where segmentation instances were removed in 7-frame-long blocks, with these elimination blocks also positioned in a uniform random manner with a 1:5 average chance of segmentation instance elimination (Box Noise 1:5). This box noise scenario presents a different but equally realistic challenge compared to the fully uniform noises, as in various applications, objects can disappear for several frames due to occlusion and limited field of view.

\begin{figure*}[]
    \centering
    \begin{subfigure}[b]{0.75\columnwidth}
        \centering
        \includegraphics[width=\linewidth]{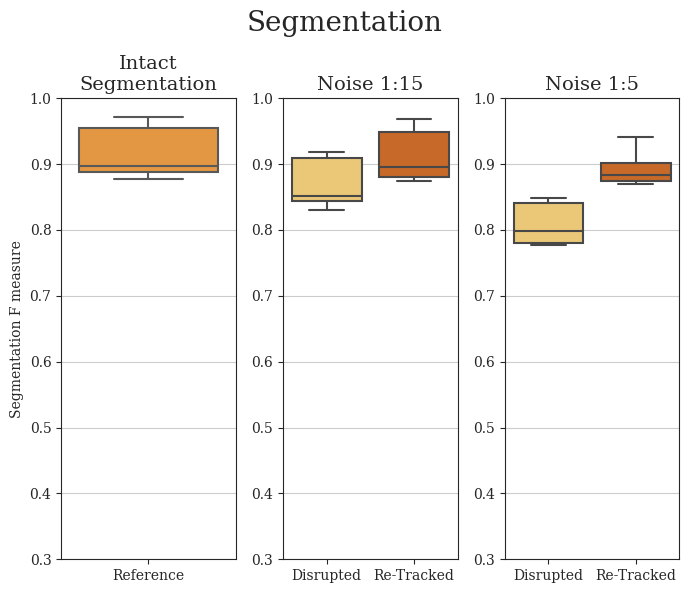}
    \end{subfigure}
     \begin{subfigure}[b]{0.75\columnwidth}
        \centering
        \includegraphics[width=\linewidth]{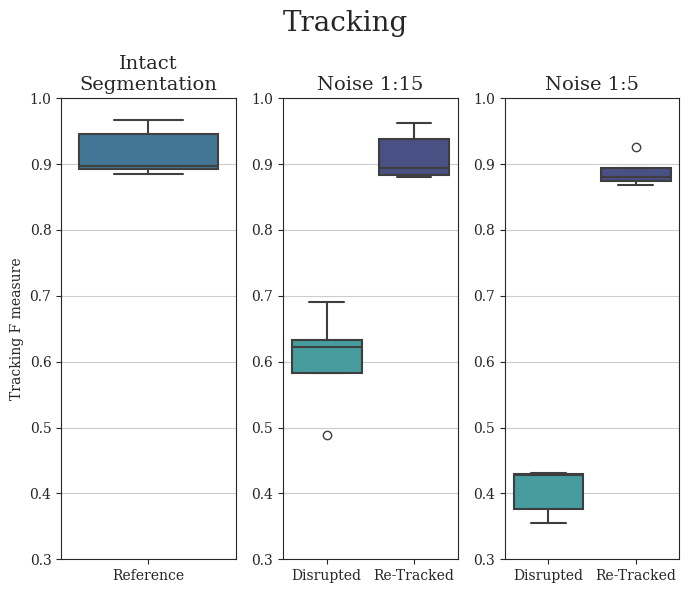}
    \end{subfigure}
    \caption{Comparative evaluation of segmentation and tracking F-scores for Intact, Noise 1:15, and Noise 1:5 cases. Both segmentation and tracking results include baseline values disrupted by the given noise, as well as re-tracked values initiated with the disrupted segmentation.}
    \label{fig:noisy_segmentation}
\end{figure*}

\begin{figure*}[]
    \centering
    \begin{subfigure}[b]{0.75\columnwidth}
        \centering
        \includegraphics[width=\linewidth]{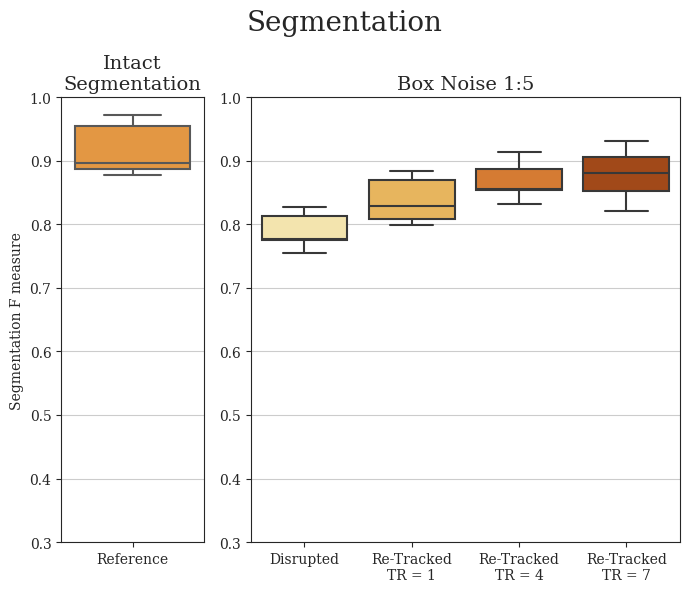}
    \end{subfigure}
     \begin{subfigure}[b]{0.75\columnwidth}
        \centering
        \includegraphics[width=\linewidth]{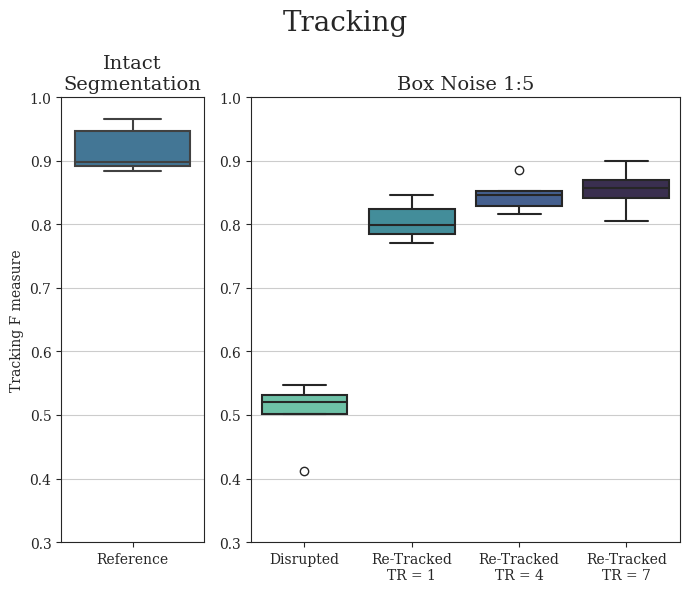}
    \end{subfigure}
    \caption{Comparative evaluation of segmentation and tracking F-scores for Intact and Box Noise 1:5 cases. Both segmentation and tracking results include baseline values disrupted by the noise, as well as re-tracked values initiated with the disrupted segmentation using local tracking ranges ($TR$) of 1, 4 and 7.}
    \label{fig:box_noisy_segmentation}
\end{figure*}

The applied noises substantially impacted both segmentation and tracking outcomes, as segmentation instances were removed and the previously continuous tracks were broken up. Therefore, it was the task of the tracker module to create continuous tracks despite the missing segmentation instances and to interpolate the removed instances as effectively as possible. The resulting F scores before and after re-tracking are presented in Figure \ref{fig:noisy_segmentation} for Noise 1:15 and Noise 1:5 with the thus far best performing local tracking range of 1. For Box Noise 1:5, we measured the performance of the architecture for local tracking ranges of 1, 4, and 7. These F scores are presented in Figure \ref{fig:box_noisy_segmentation}. 

The results reveal that for all examined noises, both the segmentation and tracking F scores showed substantial improvement compared to the disrupted tracks due to the interpolated segmentation instances and the reconnected tracks. Furthermore, in the case of Box Noise 1:5, the longer local tracking ranges were preferred, as they provided better coverage for the continuously missing segmented instances. These findings clearly demonstrate that while there is a substantial correspondence between instance segmentation and tracking results, both modules of our method strongly support each other, leading to simultaneous improvements in both, thus contributing to the observable correspondence.

\begin{figure}[]
    \centering
    \begin{subfigure}[b]{0.32\columnwidth}
        \centering
        \includegraphics[height=\linewidth]{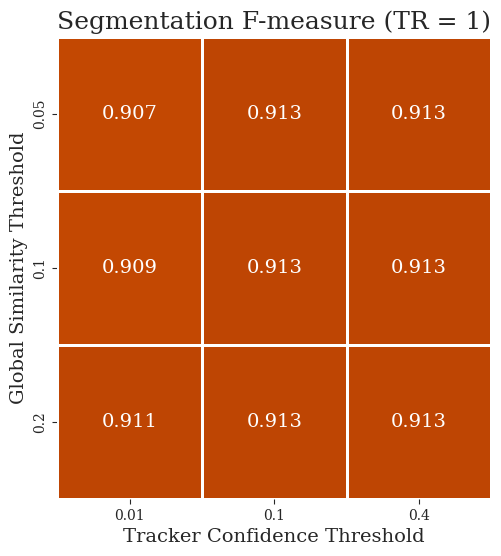}
    \end{subfigure}
    \begin{subfigure}[b]{0.32\columnwidth}
        \centering
        \includegraphics[height=\linewidth]{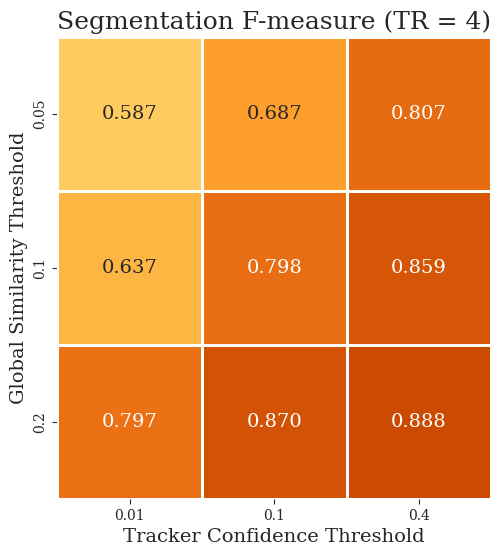}
    \end{subfigure}
    \begin{subfigure}[b]{0.32\columnwidth}
        \centering
        \includegraphics[height=\linewidth]{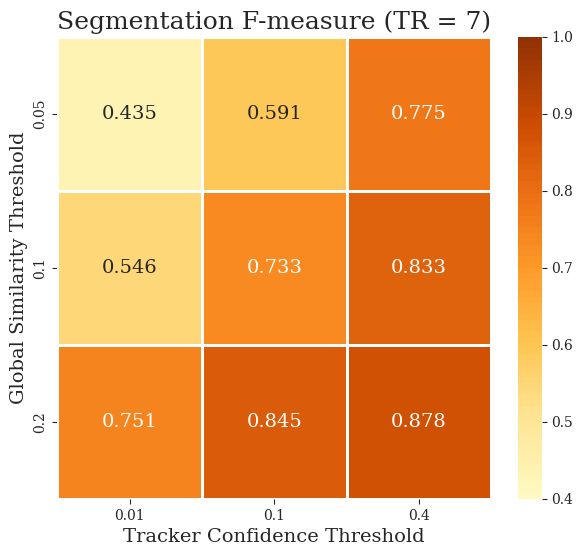}
    \end{subfigure}
    \begin{subfigure}[b]{0.32\columnwidth}
        \centering
        \includegraphics[height=\linewidth]{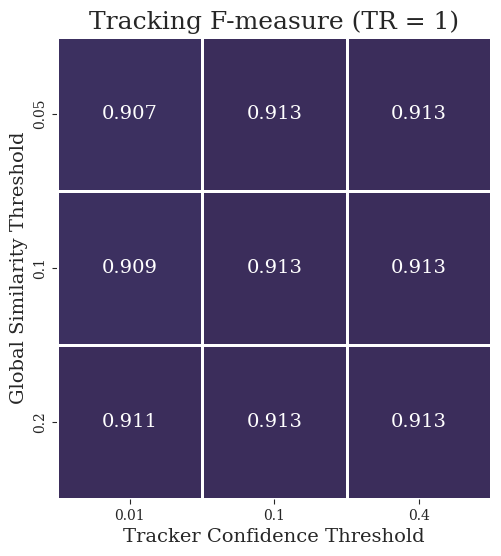}
    \end{subfigure}
    \begin{subfigure}[b]{0.32\columnwidth}
        \centering
        \includegraphics[height=\linewidth]{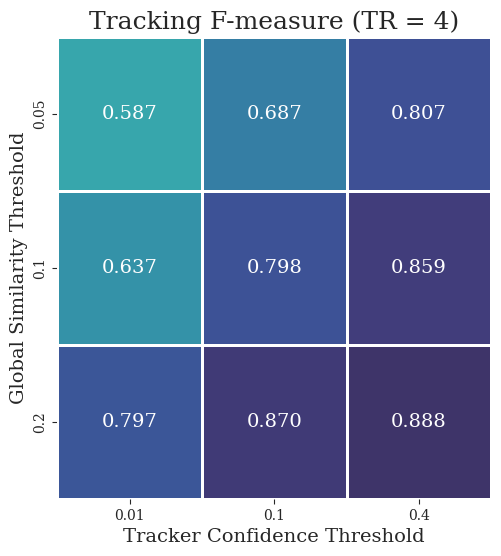}
    \end{subfigure}
    \begin{subfigure}[b]{0.32\columnwidth}
        \centering
        \includegraphics[height=\linewidth]{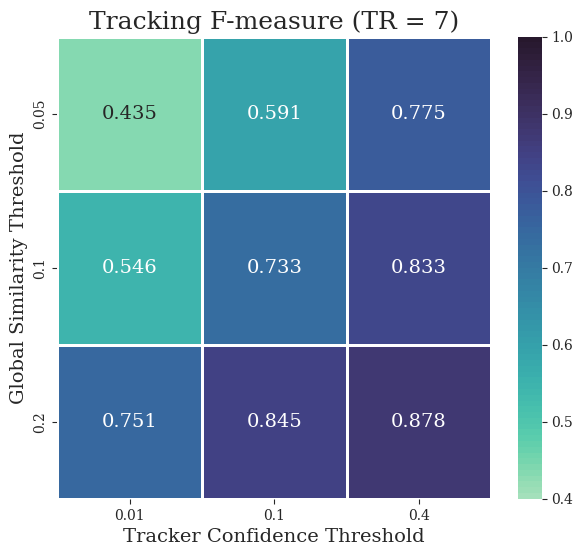}
    \end{subfigure}
    \caption{Illustration of the interdependence of hyperparameters: segmentation confidence threshold of the tracker model and minimal similarity threshold of the global assignment as a function of local tracking range ($TR$).}
    \label{fig:parameter_matrixes}
\end{figure}

Moreover, during this experiment, we observed a substantially greater influence of tracking parameters compared to previous trials. Specifically, the accuracy of the segmentation confidence threshold of the tracking model and the minimal similarity threshold of the global assignment step became notably more crucial, particularly in scenarios of reduced instance segmentation quality. This was especially evident with more complex models featuring larger local tracking ranges. An illustration of this relationship is depicted in Figure \ref{fig:parameter_matrixes} for Noise 1:15. While this observation might imply that extensive and costly hyperparameter tuning is necessary for the tracker module in case of lower quality instance segmentation results, in practice, these parameters can be readily and efficiently adjusted based on visual assessment of segmentation quality of the local tracker model and a global tracking sanity check on only a few consecutive frames. Conversely, the heightened significance of the accuracy of these parameters further underscores that while tracking may be straightforward in cases of good quality instance segmentation, achievable by nearly any model, in scenarios of poorer quality instance segmentation, a well-designed tracking architecture can play a pivotal role, substantially impacting both segmentation and tracking quality.

\subsection{Comparative Datasets}

\begin{figure*}[]
    \centering
    
    \begin{subfigure}[b]{0.4\columnwidth}
        \centering
        \includegraphics[width=\linewidth]{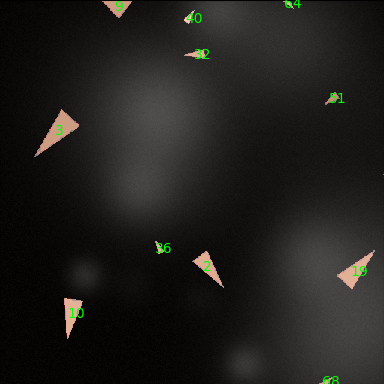}
    \end{subfigure}
    \begin{subfigure}[b]{0.4\columnwidth}
        \centering
        \includegraphics[width=\linewidth]{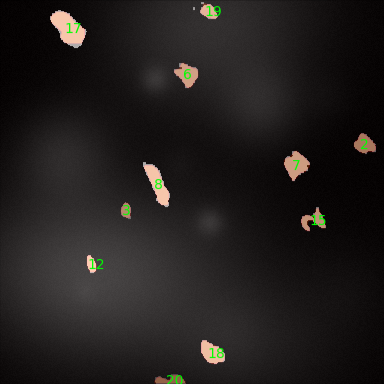}
    \end{subfigure}
    \begin{subfigure}[b]{0.4\columnwidth}
        \centering
        \includegraphics[width=\linewidth]{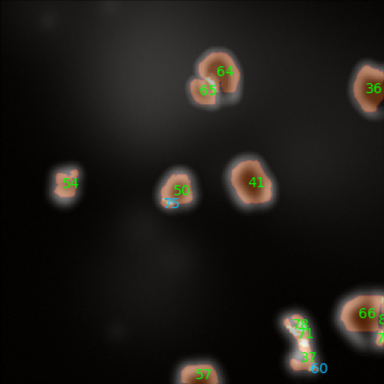}
    \end{subfigure}
    \begin{subfigure}[b]{0.4\columnwidth}
        \centering
        \includegraphics[width=\linewidth]{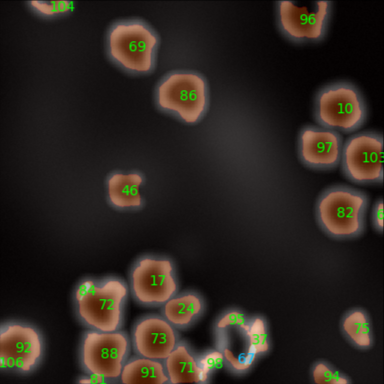}
    \end{subfigure}
    \begin{subfigure}[b]{0.4\columnwidth}
        \centering
        \includegraphics[width=\linewidth]{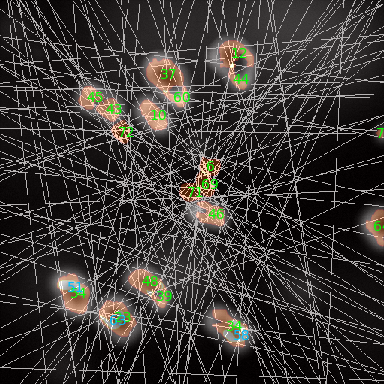}
    \end{subfigure}

    \begin{subfigure}[b]{0.4\columnwidth}
        \centering
        \includegraphics[width=\linewidth]{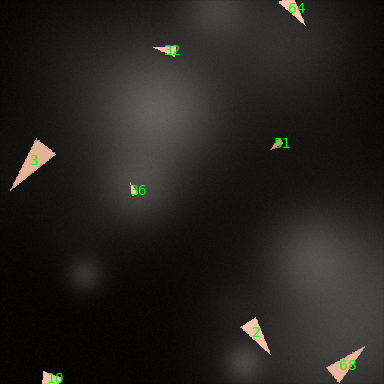}
    \end{subfigure}
    \begin{subfigure}[b]{0.4\columnwidth}
        \centering
        \includegraphics[width=\linewidth]{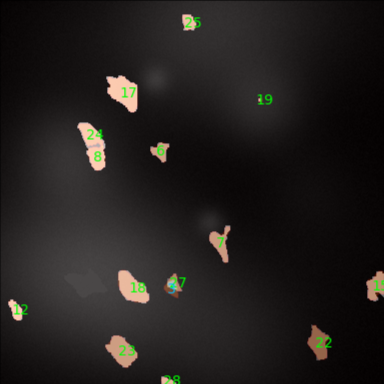}
    \end{subfigure}
    \begin{subfigure}[b]{0.4\columnwidth}
        \centering
        \includegraphics[width=\linewidth]{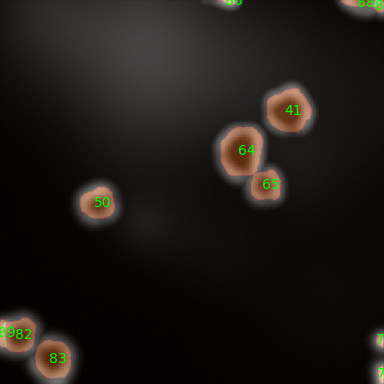}
    \end{subfigure}
    \begin{subfigure}[b]{0.4\columnwidth}
        \centering
        \includegraphics[width=\linewidth]{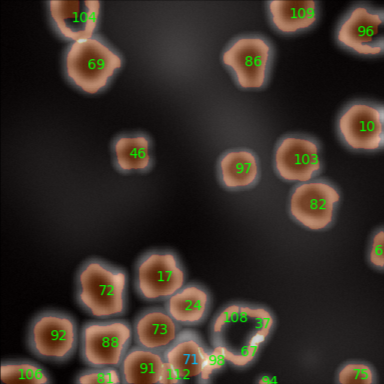}
    \end{subfigure}
    \begin{subfigure}[b]{0.4\columnwidth}
        \centering
        \includegraphics[width=\linewidth]{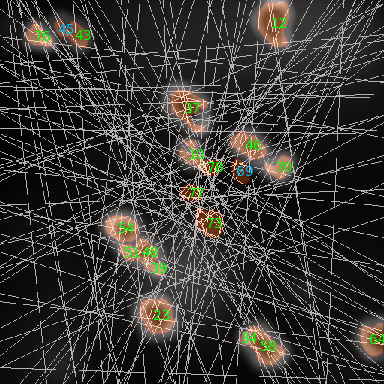}
    \end{subfigure}

    \begin{subfigure}[b]{0.4\columnwidth}
        \centering
        \includegraphics[width=\linewidth]{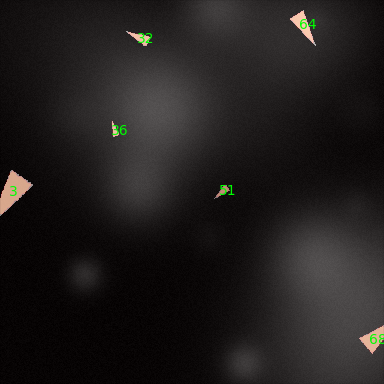}
    \end{subfigure}
    \begin{subfigure}[b]{0.4\columnwidth}
        \centering
        \includegraphics[width=\linewidth]{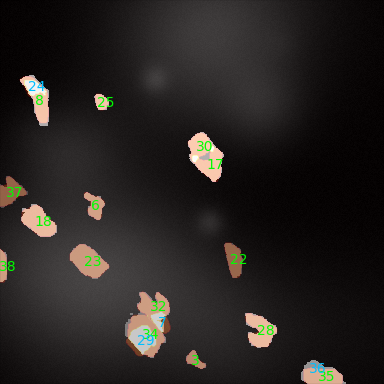}
    \end{subfigure}
    \begin{subfigure}[b]{0.4\columnwidth}
        \centering
        \includegraphics[width=\linewidth]{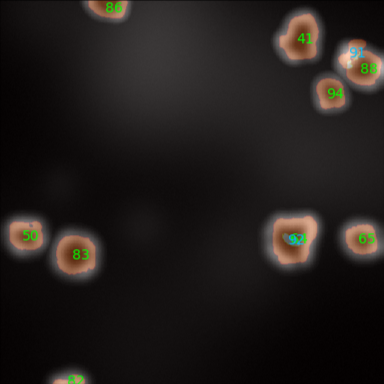}
    \end{subfigure}
    \begin{subfigure}[b]{0.4\columnwidth}
        \centering
        \includegraphics[width=\linewidth]{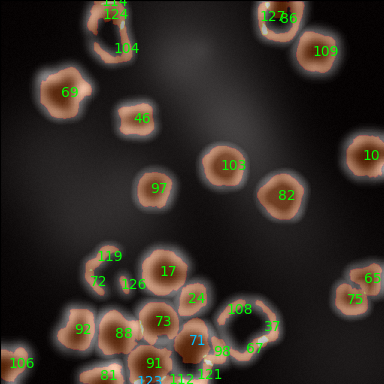}
    \end{subfigure}
    \begin{subfigure}[b]{0.4\columnwidth}
        \centering
        \includegraphics[width=\linewidth]{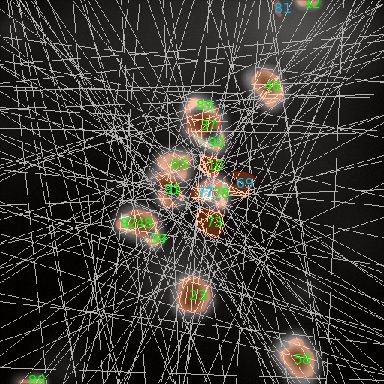}
    \end{subfigure}
    
    \caption{A display of segmentation and tracking results on various synthetic datasets with highly different object behaviors, ordered from left to right as "S. Arrows," "S. Amoeboids," "S. Amoeboids-PC," "S. Amoeboids-PCC," and "S. Amoeboids-PCCA." The displayed consecutive images are only 4 frames apart to show understandable results even in cases of extremely fast-moving objects.}
    \label{fig:synthetic_tracking_results}
\end{figure*}

Using the five synthetic datasets described in Section \ref{sec:comp_datasets_descr}, our yeast tracking architecture was trained on 80 synthesized videos for each type, each comprising 100 frames, and subsequently evaluated on 20 videos. The only modification applied to the architecture was the utilization of bounding box marking instead of centroid marking for the tracked objects. This adjustment was necessary due to the potential intersection of object paths and thus the possibility of occlusion, rendering centroid marking ambiguous in certain scenarios. The chosen $TR$ value for each training was 4, as it provides a good balance between missing instance interpolation and prediction quality. The resulting segmentation and tracking F scores are displayed in Table \ref{tab:synthetic_datasets}, while sample prediction results are displayed in \ref{fig:synthetic_tracking_results}. These sample displays were chosen in an unbiased manner by always selecting the first training sample, regardless of prediction quality. However, the frames chosen for display were selected to best showcase the differences between the datasets.

\begin{table}[]
\centering
\resizebox{\columnwidth}{!}{
\begin{tabular}{lll}
\hline
                         & Segmentation F score & Tracking F score  \\ \hline
S. Arrows         & $0.9185\pm0.0057$    & $0.8990\pm0.0076$  \\
S. Amoeboids      & $0.7137\pm0.0082$    & $0.6605\pm0.0093$ \\
S. Amoeboids-PC   & $0.6861\pm0.0089$    & $0.6662\pm0.0098$ \\
S. Amoeboids-PCC  & $0.7726\pm0.0101$    & $0.7693\pm0.0111$ \\
S. Amoeboids-PCCA & $0.5078\pm0.0129$    & $0.5022\pm0.0140$\\ \hline
Yeast Reference         & $0.9234\pm0.0136$    & $0.9202\pm0.0138$  \\ \hline
\end{tabular}
}
\caption{Segmentation and Tracking F scores of the full pipeline for various synthetic datasets with vastly different object behaviors and challenges compared to natural yeast recordings.}
\label{tab:synthetic_datasets}
\end{table}

These results clearly display the difficulty difference between objects with more predictable behaviors, such as yeast cells or synthetic arrows, and comparatively more challenging and unstable objects, such as the synthetic amoeboid versions. While the numerical results for amoeboids are lower compared to those for yeast cells or synthetic arrows, they are still acceptable given the difficulty of the tasks. The errors mostly arise from instance segmentation, as tracking F scores are never substantially lower than the segmentation F scores. Empirical assessment also shows that the predictions are mostly correct and would serve as a valuable baseline for later manual corrections if necessary. Furthermore, these results could be substantially improved through specialized augmentation techniques, hyperparameter tuning, using more complex backbone architectures for feature extraction, and increased training data. Therefore, based on our assessment, these results demonstrate the adaptability of our architecture in various object tracking scenarios with vastly different object morphologies and behaviors. However, they also highlight the different training requirements and expectations for different datasets. A more detailed analysis of this aspect is described in Section \ref{sec:data_reqs}.

\subsection{Data Requirements}\label{sec:data_reqs}

\begin{figure*}[]
    \centering
    \begin{subfigure}[b]{.55\columnwidth}
        \centering
        \includegraphics[height=\linewidth]{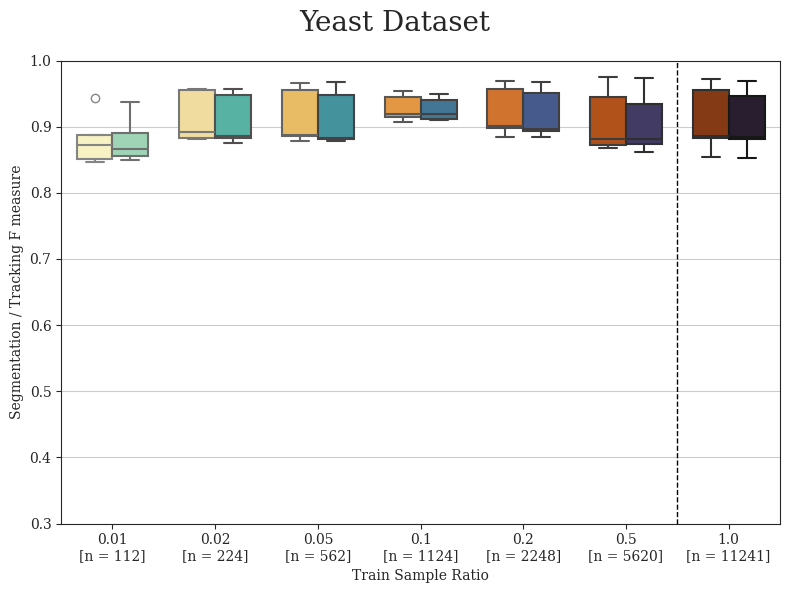}
    \end{subfigure}
    \hspace{.2\columnwidth}
    \begin{subfigure}[b]{.55\columnwidth}
        \centering
        \includegraphics[height=\linewidth]{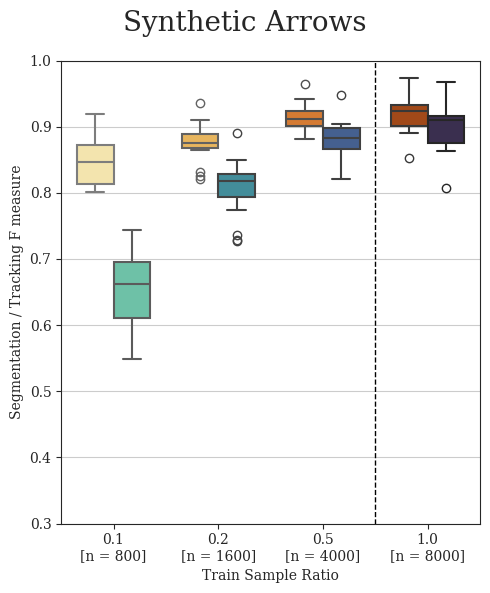}
    \end{subfigure}
    \begin{subfigure}[b]{.55\columnwidth}
        \centering
        \includegraphics[height=\linewidth]{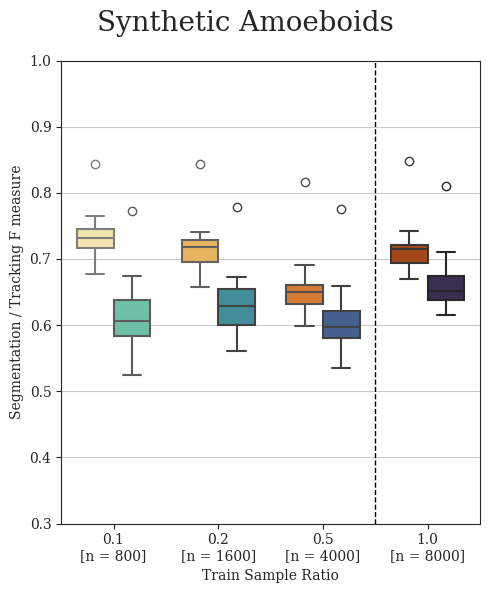}
    \end{subfigure}
    \begin{subfigure}[b]{.55\columnwidth}
        \centering
        \includegraphics[height=\linewidth]{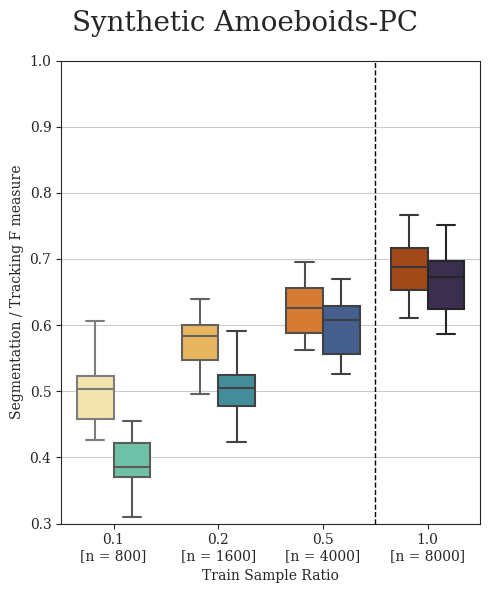}
    \end{subfigure}
    \begin{subfigure}[b]{.55\columnwidth}
        \centering
        \includegraphics[height=\linewidth]{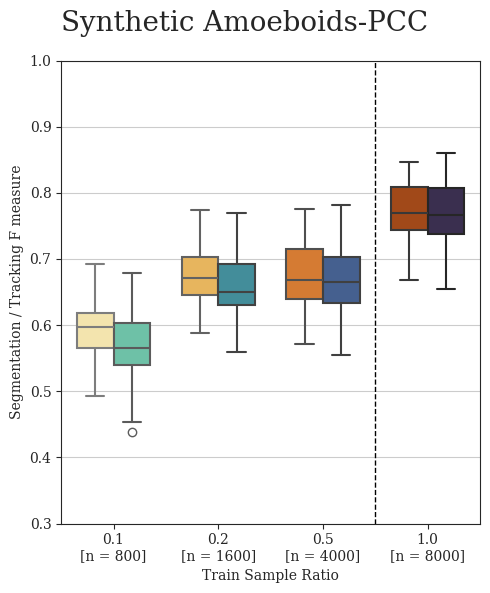}
    \end{subfigure}
    \begin{subfigure}[b]{.55\columnwidth}
        \centering
        \includegraphics[height=\linewidth]{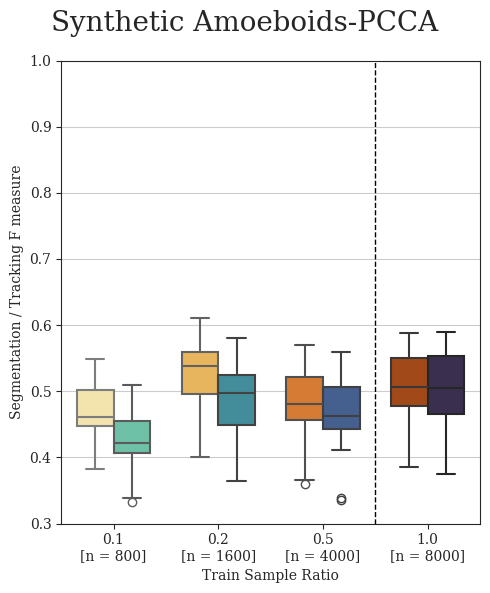}
    \end{subfigure}
    \caption{A comparison of training data requirements and the impact of reduced sample sizes on yeast cell tracking and various synthetic datasets. The left boxes show segmentation F scores, while the right boxes show tracking F scores for each individually trained model pair. The dashed lines separate results obtained using the full training set reference models from those trained on randomly selected subsets. The x-axis upper number indicates the relative training set size ratio, while "n" denotes the number of images in each set.}
    \label{fig:sr_results}
\end{figure*}

The training of neural network-based prediction models always requires a varied set of training samples, ideally covering all possible scenarios that may arise during inference. Furthermore, the number of sample points should be high enough to minimize dataset-specific learning, commonly referred to as overfitting to the training dataset. While both sample variety and the avoidance of overfitting can be improved through artificial data augmentation and other generalization techniques, there is a highly task-dependent limit for the minimal required amount of training data. Assessing this limit numerically before initial training and evaluation can be very challenging or nearly impossible. Thus, we provide example evaluation results using training sets with varying sample numbers as potential guidelines on the yeast tracking dataset, as well as the synthetic datasets described in Section \ref{sec:comp_datasets_descr}. The results using randomly selected subsets of the original dataset for the training of both instance segmentation and local tracking models are presented in Figure \ref{fig:sr_results}. For the proper interpretation of these results, it must be noted that individual model trainings can vary to a certain degree due to the stochastic nature of the training process, which is especially true for datasets with lower sample sizes. Nevertheless, some clear patterns are still noticeable. Based on these, it appears that yeast cells are by far the easiest of these objects to segment and track, as only 2\% of the 314, approximately 35.8-frame-long videos can lead to nearly identical results as training on the full dataset. Conversely, the various examined synthetic datasets showed varied results, clearly displaying how challenging the unique features of the given dataset are.

\section{Conclusion}

In this paper, we have presented a comprehensive pipeline for cell segmentation and tracking of budding yeast cells, introducing a novel architecture for cell tracking based on local time-symmetric tracking and global optimal assignment. Our approach incorporates re-trainable neural networks for both the segmentation and tracking stages, enabling the architecture's potential for generalization across different cell types, provided sufficient amount of labeled training data is available. The extensive evaluation of our method, utilizing an F-score based metric system separately for segmentation and tracking, highlights its effectiveness. Empirical results from rigorous testing demonstrate that our pipeline achieves state-of-the-art performance, outperforming the compared tools YeaZ and Phylocell in both segmentation and tracking tasks. This indicates the robustness and potential applicability of our method in cell tracking scenarios. By combining the advantages of deep learning-based segmentation and time-symmetric local tracking with linear optimal assignment based on the Hungarian method, we have developed a promising pipeline that has the potential to substantially contribute to cell tracking tasks in biological research.

\section*{Author Contributions}

GS designed and developed the primary pipeline, encompassing model design, training, and user interaction functionalities. GS played a major role in writing the manuscript. PB conceptualized the evaluation methods, generated comparative results for the assessed tools, contributed to architectural design, and participated in manuscript preparation. AC provided valuable theoretical insights and constructive feedback from a biological standpoint, contributing substantially to the manuscript. AH offered essential theoretical insights and constructive critiques from a machine learning perspective, made substantial contributions to pipeline design, and participated in manuscript preparation. All authors have reviewed and approved the final manuscript.

\section*{Acknowledgements}

PB was supported by an Umberto Veronesi fellowship. The AC lab is supported by the italian association for cancer research (AIRC IG-21556) and the Hungarian National Research Development and Innovation Office, through grant TKP2021-EGA-42. The funder played no role in pipeline design, result evaluation, or the writing of this manuscript. 
This research has also been partially supported by the Hungarian Government by the following grants: 2018-1.2.1-NKP00008: Exploring the Mathematical Foundations of Artificial Intelligence and TKP2021\_02-NVA-27 – Thematic Excellence Program.

Furthermore, we would like to express our gratitude to the anonymous reviewers, whose recommendations substantially improved the presented manuscript.

\section*{Code and Data availability}

We have made our segmentation and tracking pipeline openly accessible through a dedicated GitHub repository, which can be found at the following URL: \url{https://github.com/SzaboGergely0419/PPCU_IFOM_YeastTracker}. This repository not only hosts the main pipeline but also links to supplementary resources, including various versions of the tracker model and sample test data. Sample prediction results are uploaded directly to the repository. Furthermore, we have created a comprehensive Google Colab environment within the repository that enables the execution of the entire pipeline. Currently the training codes are not included in the public repository. However, we have made them available for review and plan to release them on the aforementioned repository in the foreseeable future.

\section*{Declaration of competing interest}

The authors declare that they have no known competing financial interests, personal relationships or non-financial competing interest that could have appeared to influence the work reported.


\begin{thebibliography}{10}
\expandafter\ifx\csname url\endcsname\relax
  \def\url#1{\texttt{#1}}\fi
\expandafter\ifx\csname urlprefix\endcsname\relax\def\urlprefix{URL }\fi
\providecommand{\bibinfo}[2]{#2}
\providecommand{\eprint}[2][]{\url{#2}}

\bibitem{SIFT}
\bibinfo{author}{Lowe, D.~G.}
\newblock \bibinfo{title}{Object recognition from local scale-invariant
  features}.
\newblock In \emph{\bibinfo{booktitle}{Proceedings of the seventh IEEE
  international conference on computer vision}}, vol.~\bibinfo{volume}{2},
  \bibinfo{pages}{1150--1157} (\bibinfo{organization}{Ieee},
  \bibinfo{year}{1999}).

\bibitem{RANSAC}
\bibinfo{author}{Fischler, M.~A.} \& \bibinfo{author}{Bolles, R.~C.}
\newblock \bibinfo{title}{Random sample consensus: a paradigm for model fitting
  with applications to image analysis and automated cartography}.
\newblock \emph{\bibinfo{journal}{Communications of the ACM}}
  \textbf{\bibinfo{volume}{24}}, \bibinfo{pages}{381--395}
  (\bibinfo{year}{1981}).

\bibitem{KalmanFilter}
\bibinfo{author}{Kálmán, R.~E.}
\newblock \bibinfo{title}{A new approach to linear filtering and prediction
  problems}.
\newblock \emph{\bibinfo{journal}{Journal of Basic Engineering}}
  \textbf{\bibinfo{volume}{82}}, \bibinfo{pages}{35--45}
  (\bibinfo{year}{1960}).

\bibitem{UnscentedKalman}
\bibinfo{author}{Julier, S.~J.} \& \bibinfo{author}{Uhlmann, J.~K.}
\newblock \bibinfo{title}{New extension of the kalman filter to nonlinear
  systems}.
\newblock In \emph{\bibinfo{booktitle}{Signal processing, sensor fusion, and
  target recognition VI}}, vol. \bibinfo{volume}{3068},
  \bibinfo{pages}{182--193} (\bibinfo{organization}{Spie},
  \bibinfo{year}{1997}).

\bibitem{hu2021celltracker}
\bibinfo{author}{Hu, T.}, \bibinfo{author}{Xu, S.}, \bibinfo{author}{Wei, L.},
  \bibinfo{author}{Zhang, X.} \& \bibinfo{author}{Wang, X.}
\newblock \bibinfo{title}{Celltracker: an automated toolbox for single-cell
  segmentation and tracking of time-lapse microscopy images}.
\newblock \emph{\bibinfo{journal}{Bioinformatics}}
  \textbf{\bibinfo{volume}{37}}, \bibinfo{pages}{285--287}
  (\bibinfo{year}{2021}).

\bibitem{tsai2019usiigaci}
\bibinfo{author}{Tsai, H.-F.}, \bibinfo{author}{Gajda, J.},
  \bibinfo{author}{Sloan, T.~F.}, \bibinfo{author}{Rares, A.} \&
  \bibinfo{author}{Shen, A.~Q.}
\newblock \bibinfo{title}{Usiigaci: Instance-aware cell tracking in stain-free
  phase contrast microscopy enabled by machine learning}.
\newblock \emph{\bibinfo{journal}{SoftwareX}} \textbf{\bibinfo{volume}{9}},
  \bibinfo{pages}{230--237} (\bibinfo{year}{2019}).

\bibitem{dietler2020convolutional}
\bibinfo{author}{Dietler, N.} \emph{et~al.}
\newblock \bibinfo{title}{A convolutional neural network segments yeast
  microscopy images with high accuracy}.
\newblock \emph{\bibinfo{journal}{Nature communications}}
  \textbf{\bibinfo{volume}{11}}, \bibinfo{pages}{5723} (\bibinfo{year}{2020}).

\bibitem{chen2021celltrack}
\bibinfo{author}{Chen, Y.} \emph{et~al.}
\newblock \bibinfo{title}{Celltrack r-cnn: A novel end-to-end deep neural
  network for cell segmentation and tracking in microscopy images}.
\newblock In \emph{\bibinfo{booktitle}{2021 IEEE 18th International Symposium
  on Biomedical Imaging (ISBI)}}, \bibinfo{pages}{779--782}
  (\bibinfo{organization}{IEEE}, \bibinfo{year}{2021}).

\bibitem{berg2019ilastik}
\bibinfo{author}{Berg, S.} \emph{et~al.}
\newblock \bibinfo{title}{Ilastik: interactive machine learning for (bio) image
  analysis}.
\newblock \emph{\bibinfo{journal}{Nature methods}}
  \textbf{\bibinfo{volume}{16}}, \bibinfo{pages}{1226--1232}
  (\bibinfo{year}{2019}).

\bibitem{vlah2022data}
\bibinfo{author}{Vlah, D.}, \bibinfo{author}{Kastrin, A.},
  \bibinfo{author}{Povh, J.} \& \bibinfo{author}{Vuka{\v{s}}inovi{\'c}, N.}
\newblock \bibinfo{title}{Data-driven engineering design: A systematic review
  using scientometric approach}.
\newblock \emph{\bibinfo{journal}{Advanced Engineering Informatics}}
  \textbf{\bibinfo{volume}{54}}, \bibinfo{pages}{101774}
  (\bibinfo{year}{2022}).

\bibitem{sun2017revisiting}
\bibinfo{author}{Sun, C.}, \bibinfo{author}{Shrivastava, A.},
  \bibinfo{author}{Singh, S.} \& \bibinfo{author}{Gupta, A.}
\newblock \bibinfo{title}{Revisiting unreasonable effectiveness of data in deep
  learning era}.
\newblock In \emph{\bibinfo{booktitle}{Proceedings of the IEEE international
  conference on computer vision}}, \bibinfo{pages}{843--852}
  (\bibinfo{year}{2017}).

\bibitem{katariya2022deeptrack}
\bibinfo{author}{Katariya, V.}, \bibinfo{author}{Baharani, M.},
  \bibinfo{author}{Morris, N.}, \bibinfo{author}{Shoghli, O.} \&
  \bibinfo{author}{Tabkhi, H.}
\newblock \bibinfo{title}{Deeptrack: Lightweight deep learning for vehicle
  trajectory prediction in highways}.
\newblock \emph{\bibinfo{journal}{IEEE Transactions on Intelligent
  Transportation Systems}} \textbf{\bibinfo{volume}{23}},
  \bibinfo{pages}{18927--18936} (\bibinfo{year}{2022}).

\bibitem{charvin2021phylocell}
\bibinfo{author}{Charvin, G.}
\newblock \bibinfo{title}{Phylocell} (\bibinfo{year}{2021}).

\bibitem{mavska2023cell}
\bibinfo{author}{Ma{\v{s}}ka, M.}, \bibinfo{author}{Ulman, V.},
  \bibinfo{author}{Delgado-Rodriguez, P.}, \bibinfo{author}{G{\'o}mez-de-Mariscal, E.},
  \bibinfo{author}{Ne{\v{c}}asov{\'a}, T.}, \bibinfo{author}{Guerrero Pe{\~n}a, F. A.},
  \bibinfo{author}{Ren, T. I.}, \bibinfo{author}{Meyerowitz, E. M.}, 
  \bibinfo{author}{Scherr, T.}, \bibinfo{author}{L{\"o}ffler, K.}, \emph{et al.}
\newblock \bibinfo{title}{The Cell Tracking Challenge: 10 years of objective benchmarking}.
\newblock \emph{\bibinfo{journal}{Nature Methods}} \textbf{\bibinfo{volume}{XX}}, \bibinfo{pages}{1--11} (\bibinfo{year}{2023}).

\bibitem{anjum2020ctmc}
\bibinfo{author}{Anjum, S.} and \bibinfo{author}{Gurari, D.}
\newblock \bibinfo{title}{CTMC: Cell tracking with mitosis detection dataset challenge}.
\newblock In \emph{\bibinfo{booktitle}{Proceedings of the IEEE/CVF Conference on Computer Vision and Pattern Recognition Workshops}}, \bibinfo{pages}{982--983} (\bibinfo{year}{2020}).

\bibitem{toubal2023ensemble}
\bibinfo{author}{Toubal, I. E.}, \bibinfo{author}{Al-Shakarji, N.}, \bibinfo{author}{Cornelison, D. D. W.}, \bibinfo{author}{Palaniappan, K.}
\newblock \bibinfo{title}{Ensemble Deep Learning Object Detection Fusion for Cell Tracking, Mitosis, and Lineage}.
\newblock \emph{\bibinfo{journal}{IEEE Open Journal of Engineering in Medicine and Biology}} (\bibinfo{year}{2023}).

\bibitem{perlin1985image}
\bibinfo{author}{Perlin, K.}
\newblock \bibinfo{title}{An image synthesizer}.
\newblock \emph{\bibinfo{journal}{ACM Siggraph Computer Graphics}} \textbf{\bibinfo{volume}{19}}, \bibinfo{number}{3}, \bibinfo{pages}{287--296} (\bibinfo{year}{1985}).

\bibitem{canny1986computational}
\bibinfo{author}{Canny, J.}
\newblock \bibinfo{title}{A computational approach to edge detection}.
\newblock \emph{\bibinfo{journal}{IEEE Transactions on Pattern Analysis and Machine Intelligence}} \bibinfo{number}{6}, \bibinfo{pages}{679--698} (\bibinfo{year}{1986}).

\bibitem{SegNet}
\bibinfo{author}{Badrinarayanan, V.}, \bibinfo{author}{Kendall, A.} \&
  \bibinfo{author}{Cipolla, R.}
\newblock \bibinfo{title}{Segnet: A deep convolutional encoder-decoder
  architecture for image segmentation}.
\newblock \emph{\bibinfo{journal}{IEEE transactions on pattern analysis and
  machine intelligence}} \textbf{\bibinfo{volume}{39}},
  \bibinfo{pages}{2481--2495} (\bibinfo{year}{2017}).

\bibitem{Unet}
\bibinfo{author}{Ronneberger, O.}, \bibinfo{author}{Fischer, P.} \&
  \bibinfo{author}{Brox, T.}
\newblock \bibinfo{title}{U-net: Convolutional networks for biomedical image
  segmentation}.
\newblock \emph{\bibinfo{journal}{Medical Image Computing and Computer-Assisted
  Intervention (MICCAI)}} \textbf{\bibinfo{volume}{9351}},
  \bibinfo{pages}{234--241} (\bibinfo{year}{2015}).

\bibitem{MaskRCNN}
\bibinfo{author}{He, K.}, \bibinfo{author}{Gkioxari, G.},
  \bibinfo{author}{Doll{\'a}r, P.} \& \bibinfo{author}{Girshick, R.}
\newblock \bibinfo{title}{Mask r-cnn}.
\newblock In \emph{\bibinfo{booktitle}{Proceedings of the IEEE international
  conference on computer vision}}, \bibinfo{pages}{2961--2969}
  (\bibinfo{year}{2017}).

\bibitem{HungarianMethod}
\bibinfo{author}{Kuhn, H.~W.}
\newblock \bibinfo{title}{The hungarian method for the assignment problem}.
\newblock \emph{\bibinfo{journal}{Naval research logistics quarterly}}
  \textbf{\bibinfo{volume}{2}}, \bibinfo{pages}{83--97} (\bibinfo{year}{1955}).

\bibitem{Detectron2}
\bibinfo{author}{Wu, Y.}, \bibinfo{author}{Kirillov, A.},
  \bibinfo{author}{Massa, F.}, \bibinfo{author}{Lo, W.-Y.} \&
  \bibinfo{author}{Girshick, R.}
\newblock \bibinfo{title}{Detectron2}.
\newblock
  \bibinfo{howpublished}{\url{https://github.com/facebookresearch/detectron2}}
  (\bibinfo{year}{2019}).

\bibitem{COCO}
\bibinfo{author}{Lin, T.-Y.} \emph{et~al.}
\newblock \bibinfo{title}{Microsoft coco: Common objects in context}.
\newblock In \emph{\bibinfo{booktitle}{Computer Vision--ECCV 2014: 13th
  European Conference, Zurich, Switzerland, September 6-12, 2014, Proceedings,
  Part V 13}}, \bibinfo{pages}{740--755} (\bibinfo{organization}{Springer},
  \bibinfo{year}{2014}).

\bibitem{buslaev2020albumentations}
\bibinfo{author}{Buslaev, A.} \emph{et~al.}
\newblock \bibinfo{title}{Albumentations: fast and flexible image
  augmentations}.
\newblock \emph{\bibinfo{journal}{Information}} \textbf{\bibinfo{volume}{11}},
  \bibinfo{pages}{125} (\bibinfo{year}{2020}).

\bibitem{wang2019learning}
\bibinfo{author}{Wang, X.}, \bibinfo{author}{Jabri, A.} \&
  \bibinfo{author}{Efros, A.~A.}
\newblock \bibinfo{title}{Learning correspondence from the cycle-consistency of
  time}.
\newblock In \emph{\bibinfo{booktitle}{Proceedings of the IEEE/CVF Conference
  on Computer Vision and Pattern Recognition}}, \bibinfo{pages}{2566--2576}
  (\bibinfo{year}{2019}).

\bibitem{deeplabv3plus2018}
\bibinfo{author}{Chen, L.-C.}, \bibinfo{author}{Zhu, Y.},
  \bibinfo{author}{Papandreou, G.}, \bibinfo{author}{Schroff, F.} \&
  \bibinfo{author}{Adam, H.}
\newblock \bibinfo{title}{Encoder-decoder with atrous separable convolution for
  semantic image segmentation}.
\newblock In \emph{\bibinfo{booktitle}{ECCV}} (\bibinfo{year}{2018}).

\bibitem{Iakubovskii:2019}
\bibinfo{author}{Iakubovskii, P.}
\newblock \bibinfo{title}{Segmentation models pytorch}.
\newblock
  \bibinfo{howpublished}{\url{https://github.com/qubvel/segmentation_models.pytorch}}
  (\bibinfo{year}{2019}).

\bibitem{torchvision}
\bibinfo{author}{{PyTorch Contributors}}.
\newblock \bibinfo{title}{{torchvision.transforms}}.
\newblock
  \bibinfo{howpublished}{\url{https://pytorch.org/docs/stable/torchvision/transforms.html}}
  (\bibinfo{year}{2023}).

\bibitem{arbellebgu}
\bibinfo{author}{Arbelle, A.}, \bibinfo{author}{Cohen, S.}, \bibinfo{author}{Raviv, T. R.}, \bibinfo{author}{Ben-Haim, T.}
\newblock \bibinfo{title}{BGU-IL (5) DESCRIPTION}.

\bibitem{primet2011probabilistic}
\bibinfo{author}{Primet, M.}
\newblock \emph{\bibinfo{title}{Probabilistic methods for point tracking and
  biological image analysis}}.
\newblock Ph.D. thesis, \bibinfo{school}{Universit{\'e} Ren{\'e}
  Descartes-Paris V} (\bibinfo{year}{2011}).

\bibitem{fehrmann2013aging}
\bibinfo{author}{Fehrmann, S.} \emph{et~al.}
\newblock \bibinfo{title}{Aging yeast cells undergo a sharp entry into
  senescence unrelated to the loss of mitochondrial membrane potential}.
\newblock \emph{\bibinfo{journal}{Cell reports}} \textbf{\bibinfo{volume}{5}},
  \bibinfo{pages}{1589--1599} (\bibinfo{year}{2013}).

\bibitem{padovani2022segmentation}
\bibinfo{author}{Padovani, F.}, \bibinfo{author}{Mairh{\"o}rmann, B.},
  \bibinfo{author}{Falter-Braun, P.}, \bibinfo{author}{Lengefeld, J.} \&
  \bibinfo{author}{Schmoller, K.~M.}
\newblock \bibinfo{title}{Segmentation, tracking and cell cycle analysis of
  live-cell imaging data with cell-acdc}.
\newblock \emph{\bibinfo{journal}{BMC biology}} \textbf{\bibinfo{volume}{20}},
  \bibinfo{pages}{174} (\bibinfo{year}{2022}).

\end{thebibliography}

\end{document}